\documentclass[pra,amsmath,amssymb,twocolumn]{revtex4}
\usepackage[normalem]{ulem}
\usepackage{amsmath,amssymb,times}
\usepackage{graphicx}
\usepackage[usenames]{color}

\newcommand{\beq}{\begin{eqnarray}}
\newcommand{\eeq}{\end{eqnarray}}
\newcommand{\ms}{{m_s}}

\def \ra{{\rightarrow}}
\def \ua{{\uparrow}}
\def \da{{\downarrow}}

\def \half{\frac{1}{2}}
\def \etal{{\it {et al.}}}

\def \e{{\epsilon}}

\def \L{{\Lambda}}
\def \a{{\alpha}}

\def \b{{\beta}}
\def \g{{\gamma}}
\def \D{{\Delta}}
\def \d{{\delta}}
\def \w{{\omega}}
\def \s{{\sigma}}

\def \e{{\epsilon}}

\def \br{{\bf r}}

\def \tone{{\tau_1}}
\def \ttwo{{\tau_2}}

\def \av#1{{\langle#1\rangle}}
\def \ket#1{{|#1\rangle}}
\def \bra#1{{\langle #1|}}
\def \braket#1#2{{\langle #1|#2 \rangle }}

\newcommand{\gap}{\Delta_{s}}
\newcommand{\noss}{1}
\newcommand{\oss}{2}

\def \figwa{0.44}
\def \figwb{0.34}
\def \figw{0.36}
\def \figw3{0.24}
\def \figpreamble{}

\newcommand{\fribourg}{Department of Physics, University of Fribourg, CH-1700 Fribourg, Switzerland}
\newcommand{\harvard}{Department of Physics, Harvard University, Cambridge, MA 02138}

\begin{document}

\title{Quantum quenches in the anisotropic spin-1/2 Heisenberg chain: different approaches to many-body dynamics far from equilibrium}

\author{Peter Barmettler}
\affiliation{\fribourg}
\author{Matthias Punk}
\affiliation{Department of Physics, Technical University Munich, D-85748 Garching, Germany}
\author{Vladimir Gritsev}
\affiliation{\fribourg}
\author{Eugene Demler}
\affiliation{\harvard}
\author{Ehud Altman}
\affiliation{Department of Condensed Matter Physics, Weizmann Institute of Science, Rehovot, 76100, Israel}

\date{November 10, 2009}

\begin{abstract}
Recent experimental achievements in controlling ultracold gases in optical lattices  open a new perspective on quantum many-body physics. In these experimental setups it is possible to study coherent time evolution of isolated quantum systems. These dynamics reveal new physics beyond the low-energy properties usually relevant in solid-state many-body systems. In this paper we study the time evolution of antiferromagnetic order in the Heisenberg chain after a sudden change of the anisotropy parameter, using various numerical and analytical methods. As a generic result we find that the order parameter, which can show oscillatory or non-oscillatory dynamics, decays exponentially except for the effectively non-interacting case of the XX limit.
For weakly ordered initial states we also find evidence for an algebraic correction to the exponential law.
The study is based on numerical simulations using a numerical matrix product method for infinite system sizes (iMPS), for which we provide a detailed description and an error analysis. Additionally, we investigate in detail the exactly solvable XX limit. These results are compared to approximative analytical approaches including an effective description by the XZ-model as well as by mean-field, Luttinger-liquid and sine-Gordon theories. This reveals which aspects of non-equilibrium dynamics can as in equilibrium be described by low-energy theories and which are the novel phenomena specific to quantum quench dynamics. The relevance of the energetically high part of the spectrum is illustrated by means of a full numerical diagonalization of the Hamiltonian.

\end{abstract}

\maketitle
\tableofcontents
\clearpage
\section{Introduction}
\subsection{Motivation}

Developing a profound understanding of interacting quantum many-body systems is one of the important challenges in modern physics. Potential applications of correlated quantum systems have driven decades of theoretical and experimental investigation. There are very well understood classes of many-body systems which essentially behave as ensembles of non-interacting particles, with Landau's Fermi liquid theory as the most prominent example. This picture can however break down in the presence of strong correlations or in reduced dimensions. 
For high-temperature superconductivity \cite{bednorz-1986}, in the quantum-Hall effect or for transport in semiconductor nanodevices interactions lead to intricate quantum many-body phenomena for which the theoretical basis is still incomplete.

A variety of analytical and numerical techniques have been developed in order to study models of correlated systems. However, the degrees of freedom in quantum mechanics grow in general exponentially with the number of particles -- a fundamental problem which can make the analytical approach highly complex and restricts the applicability of computer simulations. An alternative to the analytical and numerical treatment of the quantum many-body problem has been proposed by R. Feynman \cite{feynman-1986}, who introduced the idea of \textit{quantum simulation}: instead of solving the highly complex theory on a computer, the model could be tested directly by means of an artificially engineered quantum system. Over the last decade, remarkable experimental setups have been developed to produce and control dilute ultracold atomic gases in optical lattices \cite{bloch-2008}. These systems are very promising candidates for the realization of Feynman's idea of quantum simulation. The key to reach collective quantum states of atomic gases was the development of laser and evaporative cooling techniques, which allow to go down to temperatures in the nanokelvin range and led to the first realizations of Bose-Einstein condensation in 1995 \cite{anderson-1995,bradley-1995,davis-1995}.
Subsequently, using optical lattices, it became possible to drive ultracold gases into bosonic  \cite{greiner-2001a,greiner-2001} and fermionic \cite{chin-2006,zwierlein-2006,jordens-2008} correlated states.

A particularity of ultracold atomic systems is the controllability of internal parameters, which relies on the development of magnetic and optical traps of various geometries \cite{bloch-2008}, and the tuning of atom-atom interactions by means of Feshbach resonance in an external magnetic field \cite{feshbach-1958,tiesinga-1993}. These parameters can be changed quickly, producing a so-called \textit{quantum quench}, which allows to generate various types of non-equilibrium situations \cite{greiner-2002,rabl-2003,kinoshita-2006,hofferberth-2007,lee-2007,trotzky-2008,palzer-2009}.  
Unlike solids, where the electronic system suffers from dissipation due to the coupling to lattice phonons, atomic gases are almost perfectly isolated quantum systems and exhibit coherent dynamics over large periods of time.
Since the coherent dynamical processes are determined exclusively by the intrinsic properties of system, it allows to investigate specifically the non-linear interaction effects. This is a unique situation, not available in usual solid-state experiments, where dynamical effects beyond linear response are in general intricate.
The theoretical study of the non-perturbative many-body aspects of non-equilibrium dynamics is the main topic of this paper. We namely focus on quantum spin chains, for which corresponding experiments are currently under development \cite{trotzky-2008,trotzky-private}. 

Due to weak dissipation in ultracold atomic gases non-equilibrium properties are important even if equilibrium aspects of these systems shall be investigated. When attempting to use ultracold gases as quantum simulators for certain equilibrium problem, one usually prepares the system in an uncorrelated initial state, e.g. a Bose-Einstein condensate, and drives it into a correlated state by a slow change of parameter \cite{greiner-2001a,trotzky-preprint-2009}. However, it has been found that in the vicinity of a phase transition the time scales needed for a correlated state to equilibrate can become exceedingly large \cite{damski-2005,dziarmaga-2005,cherng-2005,pellegrini-2008,polkovnikov-gritsev-2008,zakrzewski-2009}. It is therefore mandatory to establish relations between equilibrium and non-equilibrium systems to overcome this problem.

The relevance of non-equilibrium dynamics of cold atoms goes beyond the investigation of fundamental questions of quantum statistical mechanics. There are possible practical applications in quantum metrology \cite{giovannetti-2006} and quantum information processing \cite{mandel-2003,mandel-2003a,widera-2005,folling-2007} and relations to open questions in cosmology \cite{kribble-1976,zurek-1985}. However, before the ultracold quantum gases can be routinely applied in the context of such problems, a number of open experimental challenges need to be solved. By providing exact results for realistic experimental setups in this article we intend to support ongoing efforts in improving the control of ultracold atomic gases.

We will study the emerging dynamics of the order parameter of a XXZ Heisenberg chain prepared in the classical (uncorrelated) N\'eel state, which can be realized in experiment, but, in order to get a deeper insight into the problem, general antiferromagnetic initial states are also considered. Our special interest concerns  the effect of the quantum phase transition  crossed when tuning the magnetic anisotropy parameter.

Exact results based on numerical calculations are presented. Furthermore, alternative approximative approaches are applied. The applicability of the analytical tools, which have been very successful in the description of equilibrium phenomena, turns out to be strongly restricted for the non-equilibrium problem under consideration. We identify the apparent problems in the standard approximations and point out in which direction these approaches should be extended in order to capture the main features of the quantum quench dynamics.

\subsection{Brief review on non-equilibrium dynamics}
Mostly in relation to transport phenomena, non-equilibrium problems have been subject to intensive theoretical investigation over many years (e.g. Ref. \cite{zotos-1999}). However, non-equilibrium transport, which can be seen as a result of perturbations (voltage biases) at the edges of the system, is fundamentally different from quench dynamics, where the parameter change is global. More closely related to a quantum quench are highly excited  electronic states in solids, generated in femtosecond pump-probe spectroscopy~\cite{ogasawara-2000,iwai-2003,perfetti-2006}.
Nevertheless, in these systems decoherence times are short and the dissipative processes strongly contribute to the emerging dynamics. 
Consequently, concepts developed for transport phenomena and dynamics in condensed matter systems are not necessarily appropriate to quenches in ultracold atomic systems. Except for pioneering works on quench dynamics in the 1970's \cite{pfeuty-1970,barouch-1970a,barouch-1971b,barouch-1971c}, specific theoretical research on quench dynamics has only started in recent years, stimulated by the experimental developments in ultracold atomic physics.
In these works, which shall be briefly summarized in this section, two main lines have been followed. A first aspect is the study of the nature of the quasi-stationary states in the long-time limit. As demonstrated in an experiment of Kinoshita \etal~\cite{kinoshita-2006}, these non-equilibrium states can exhibit striking properties for specific types of interactions. Another approach explicitly focuses on the characteristics of the time evolution after the quench -- experimental examples are the oscillations \cite{greiner-2002} or the dephasing \cite{hofferberth-2007} of the superfluid phase. It turns out to be an ambitious challenge to establish relations between dynamical phenomena and the details of the microscopic model, such as integrability and dimensionality. Although numerous remarkable theoretical efforts revealed a number of interesting phenomena, many aspects of relaxation dynamics and equilibration, which shall be discussed in detail in this work, remain unclear.

The effective description of many-body systems by means of low-energy theories, captured within the renormalization group framework \cite{shankar-1994}, has proven sufficient for the theoretical understanding of a broad range of equilibrium phenomena.  Therefore the application of renormalization group ideas to non-equilibrium dynamics seems to be a promising approach. Along this way diagrammatic techniques \cite{rey-2004,gasenzer-2006,gasenzer-2007,gasenzer-2009} and the solutions of the dynamics of field-theoretical models at the renormalization group fixed point \cite{berges-2004,bettelheim-2006,cazalilla-2006,calabrese-2006,calabrese-2007,burkov-2007,gritsev-2007,iucci-2009,sabio-preprint-2009} were developed. Providing a generic view on the quench problem for critical theories, the work of Calabrese and Cardy \cite{calabrese-2006,calabrese-2007}  based on conformal field theory has to be emphasized. While for continuum systems field-theoretical models were successfully applied to generic quantum quenches \cite{burkov-2007,hofferberth-2007}, it has to be clarified under what conditions  they all provide accurate decription of lattice systems. Unclear is also the range of applicability of semiclassical theories \cite{polkovnikov-2002,polkovnikov-2003,polkovnikov-2009}.

For a restricted class of problems the time-evolution can be calculated exactly, e.g. for Jordan-Wigner diagonalizable XY-chains \cite{pfeuty-1970,barouch-1970a,barouch-1971b,barouch-1971c,igloi-2000,sengupta-2004,rigol-2007,gangardt-2008} or the $\frac{1}{r}$-Hubbard-chain \cite{kollar-2008}. A major drawback of these exactly solvable models is that the possibility of their representation in terms of non-interacting particles apparently leads to very specific relaxation phenomena, which are not generic not only for non-integrable, but also for more complicated integrable models. For instance, it is questionable whether the generalized Gibbs ensemble, which has been proposed for the description of quasi-stationary states of integrable models \cite{rigol-2007}, is a useful concept beyond the simple Jordan-Wigner diagonalizable cases \cite{gangardt-2008,kollath-2009}. For the more general Bethe-ansatz solvable models, it has not yet been possible to extract dynamics, except for the Richardson \cite{faribault-2009} and the Lieb-Liniger models \cite{gritsev-2009}.

In view of the high complexity of the quench dynamics, efficient unbiased numerical approaches are crucial to gain deeper insight. Using exact diagonalization \cite{rigol-2008,lauchli-2008,roux-2009} it is possible to calculate the dynamics of small systems over exceedingly long times. For larger (but one-dimensional) systems the density matrix renormalization group (DMRG) \cite{vidal-2003,daley-2004,white-2004,schollwoeck-2005} can be applied. Although only for finite times, dynamics of spin-chains (respectively spinless fermions) \cite{gobert-2005,chiara-2006,manmana-2007,barmettler-2008,barthel-2008,manmana-2009,barmettler-2009} and bosonic lattice models \cite{lauchli-2008,kollath-2007,cramer-2008,flesch-2008} have been evaluated. Recently, the dynamical mean field theory has been applied to fermionic models in the limit of infinite dimensions \cite{freericks-2006,eckstein-2008,hackl-2009,eckstein-preprint-2009}.

\subsection{Basic setup and general discussion}
The Heisenberg model is a paradigm in the theory of magnetism and strongly correlated systems in general. In appendix \ref{sec:magnetismexp} we derive how the model can be realized with ultracold two-level atoms in various geometries of optical lattices. For instance, it is possible generate a one-dimensional XXZ Heisenberg model,
\begin{equation}
H = J \sum_j \left\{ S_j^x S_{j+1}^x + S_j^y S_{j+1}^y + \Delta  S_j^z S_{j+1}^z \right\}\,,
\label{eq:xxzhamiltonianoned}
\end{equation}
where the sign and the strength of the exchange coupling $J$ and $\D$ can be tuned dynamically.  
The XXZ model is integrable and its eigenstates can be constructed by the Bethe ansatz. In the case of antiferromagnetic couplings $J>0$, the anisotropy parameter triggers a quantum phase transition from a gapless "Luttinger liquid" phase ($0\leq\D<1$) to a gapped, Ising-ordered antiferromagnetic phase ($\D>1$). The main features of the model at equilibrium and its field-theoretical formulation are given in appendix \ref{sec:magnetismtheo}.

The non-equilibrium dynamics (\ref{eq:xxzhamiltonianoned}) shall be investigated in the following quantum quench:  At time $t<0$ the system is prepared in a ground state $\ket{\psi_0}$ with long-range antiferromagnetic order. The corresponding anisotropy parameter is denoted as $\D_0$, $\D_0>1$. Among the aniferromagnetic equilibrium states the N\'eel state,
\beq
|\psi\rangle_{\text{N\'eel}} = | \ua \da \ua \dots \da \ua \da \rangle\,,
\label{eq:neel}
\eeq
which corresponds to the limit $\D_0\ra\infty$, has already been realized in experiment \cite{trotzky-2008} and will attract our special attention. At $t=0$ the system is pushed out of equilibrium by changing the strength of the interaction, $\D<\D_0$, and the dynamics emerging  at $t>0$ are studied. 

In the context of optical lattices, where the system is well isolated and no phonons are present, dissipation can be neglected in a first approximation. Also, being interested in quantum effects, we set $T=0$. Finite temperature may become relevant for weak magnetic exchange interactions in the ultracold atomic setup, but how to investigate efficiently the non-equilibrium problem at $T>0$ is still an unsolved problem. Under these assumptions, the dynamics is formally described by the solution of the Schr\"odinger equation,
\beq
|\psi (t)\rangle=e^{-iHt}|\psi_0\rangle\,.
\label{eq:timeevolution}
\eeq
we set $\hbar=1$ throughout this paper. Involving a priori all the energy scales of the many-body Hamiltonian, the calculation of the time evolution of the wave function (\ref{eq:timeevolution}) is highly complex. When approaching the problem analytically, one is forced to intoduce an appropriate approximation -- the advantages and drawbacks of various approaches will be investigated in this work. When using numerics the dynamics (\ref{eq:timeevolution}) can be solved by fully diagonalizing the Hamiltonian $H$. In section \ref{sec:spectral} we apply the full numerical diagonalization approach. Highly efficient routines have been developed for this purpose \cite{lapack-1999}, which can nevertheless  be used only for small system sizes (up to 20 lattice sites).
More efficient and applicable directly in the thermodynamic limit are matrix product states (MPS), which will be used for the simulation of the general quench dynamics of the XXZ model. For a detailed description of the MPS method see appendix \ref{sec:mps}.

To describe the dynamics of the state $\ket{\psi(t)}$ we mainly focus on the antiferromagnetic order parameter,
\beq
\ms(t)=\frac{1}{N}\sum_j(-1)^{j}\av{S_j^z(t)}\,.
\eeq
Since the state $\ket{\psi_0}$ is invariant under translation and subsequent spin-inversion, $\pm\ms(t)$ corresponds to the local magnetization at any site of the lattice. It will also be useful to look at the frequency distribution $f_{\ms}(\e)$, which resolves the contributions to the dynamics in energy space,
\beq
\ms(t)=\int d\e e^{-i\e t} f_{\ms}(\e)\,.
\label{eq:energydist}
\eeq
The staggered magnetization is not only the natural observable characterizing the ordering of antiferromagnetic states, but also reflects the properties of the local density matrix of a single site.
For describing non-local properties we choose the equal-time connected spin-spin correlation function,
\beq
G_c^{zz}(\ell,t)=\frac{1}{N}\sum_i\langle S^z_{i+\ell}(t)S^z_{i}(t) \rangle - \langle S^z_{i+\ell}(t)\rangle \langle S^z_{i}(t) \rangle\,.
\label{eq:spinspin}
\eeq

Before going into the study of the many-body dynamics of the Hamiltonian (\ref{eq:xxzhamiltonianoned}), it is worthwhile considering the case of only 2 spins. A corresponding experiment has been carried out by Trotzky \etal ~\cite{trotzky-2008} by loading  $^{87}$Rb atoms in the hyperfine states $
\ket{\downarrow}=|F=1,m_F=-1\rangle\,,
\ket{\uparrow}=|F=1,m_F=1\rangle\,,$
into an array of double-well potentials. The initial N\'eel state was generated using a magnetic field gradient transferring  the effective spins in each double well from a triplet-bond state into an antiferromagnetic one, $\ket{\psi_0}=\ket{\ua\da}$. The dynamics are in this special case independent of $\D$ and can be described as Rabi oscillations between $\ket{\ua\da}$ and $\ket{\da\ua}$ states, 
\beq
\ket{\psi(t)}=\cos(Jt/2)\ket{\ua\da}+i\sin(Jt/2)\ket{\da\ua}\,.
\label{eq:rabi}
\eeq
Hence, the antiferromagnetic order parameter descibes an oscillatory behaviour, $
\ms(t)=\frac{1}{2}\cos(Jt)$, where the Rabi frequency is set by the exchange coupling $J$, which was indeed observed in the experiment \cite{trotzky-2008}.

Although, as we shall see in section \ref{sec:xx}, in a many-body system such Rabi-like oscillations may survive, the dynamics become much more intricate when going to large system sizes. On a heuristic level the initial state may be regarded as a bunch of excitations of the Hamiltonian $H$, whose dynamics gives rise to the propagation of correlations throughout the system. For spin models with sufficiently local interactions, Lieb and Robinson \cite{lieb-1972} have proven that this propagation takes place within a light-cone -- the deviation of a correlation function from its initial value becomes exponentially small for distances $\ell>2ut$, where $u$ is the maximum velocity of excitations in the system. For an isolated but arbitrarily large system this means that relaxation to a stationary state can only be observed for subsystems of size $\ell<2ut$. This light-cone effect has been more precisely described in the framework of boundary conformal field theory \cite{calabrese-2005}, which predicts an exponential decay of the correlations in the long-time limit. These short-range correlations are in contrast with the entanglement properties of the non-equilibrium problem. It has been shown \cite{calabrese-2005} that the entanglement entropy of a subsystem of size $\ell$ grows linearly with time if $2ut<\ell$ and saturates to a value proportional to $\ell$ if $2ut>\ell$. 

It is an open question, under what conditions the stationary state in the long-time limit can be described by a statistical ensemble at a finite temperature, meaning that \textit{thermalization} occurs. There are several examples for which this is not the case \cite{kollath-2007,rigol-2007,manmana-2007,barmettler-2008,roux-2009}, and the extended Gibbs ensemble \cite{rigol-2007}, which takes into account the constraints of the non-dissipative dynamics, or the micro-canonical ensemble \cite{rigol-2008,deutsch-1991,kollath-2009} are possible candidates for describing steady states. Whether the integrability is a necessary condition for the absence of thermalization remains unclear. It has been pointed out that the absence of thermalization can be associated with a non-perturbative behavior, which is not related to the integrability of the underlying Hamiltonian \cite{roux-2009}.

\begin{table*}[ht]
\caption{\label{tab:results}Exact analytical and numerical results for the quench in the XXZ model. See sections \ref{sec:xx} - \ref{sec:ll} for details.}
\begin{ruledtabular}
\begin{tabular}{lccc}
initial &  coupling & asymptotic law & relaxation times\\
state 	&&& (if applicable) \\
\hline
\hline
\multicolumn{3}{l}{Sec. \ref{sec:xx}: Exact analytical calculation in the XX limit}\\
\hline
N\'eel& $\D=0$& $t^{-\frac{1}{2}} \cos(2 J t - \frac{\pi}{4})$&$\tone\approx0\,,\,\,\ttwo\rightarrow \infty$\\
SDW & $\D=0$& $ \sqrt{\frac{\Delta_s}{J}}t^{-\frac{1}{2}} e^{-2 \Delta_s t} + \frac{\Delta_s}{J}t^{-\frac{1}{2}}\cos(2Jt-\frac{\pi}{4})$& $\tone=\frac{1}{2\Delta_s}\,,\,\,\ttwo\rightarrow \infty$\\
\hline
\hline
\multicolumn{3}{l}{Sec. \ref{sec:xxz}: Numerical iMPS calculation of the XXZ model}\\
\hline
N\'eel & $\D\gtrsim 0$  & $e^{-t/\ttwo}\cos(\w t +\phi)\,$\footnotemark[2]&$\tone\approx0\,,\,\,\ttwo\sim \log\D$\\
N\'eel & $\D\gg1$  & $e^{-t/\tone}$&$\tone\sim \D^{2}$\\
$\D_0\gtrsim1$& $\D=0$ &$C_1  t^{-\frac{1}{2}}e^{-t/\tau_{\noss}} +C_2t^{-\frac{1}{2}}\cos(\omega t +\phi)\,$\footnotemark[2]&$\tone\sim\frac{1}{2\gap}\footnotemark[2]\,,\ttwo\ra\infty$\\
$\D_0\gtrsim1$& $\D\gtrsim 0$ &$C_1  t^{-\frac{1}{2}}e^{-t/\tau_{\noss}} +C_2 e^{-t/\tau_{\oss}}\cos(\omega t +\phi)\,$\footnotemark[2]&$\tone\sim\frac{1}{K\gap}\footnotemark[3]\,,\ttwo\sim \log\D$\\
$\D_0\gg1$ & $\D_0>\D\gg1$  & $e^{-t/\tone}$&$\tone\sim |\frac{1}{\Delta}-\frac{1}{\D_0}|^{-2}$\\
\hline
\hline
\multicolumn{3}{l}{Sec. \ref{sec:meanfield}: Mean field theory}\\
\hline
N\'eel&  $1>\D>0$ & $t^{-\frac{2}{3}}\left\{C_1 \cos(\w_1t+ \phi_1) + C_2 \cos(\w_2t+ \phi_2)\right\}$&$\tone\approx0\,,\,\,\ttwo\rightarrow \infty$\\
N\'eel&  $\D>1$ & $ const.$&\\
\hline
\hline
\multicolumn{3}{l}{Sec. \ref{sec:xz}: XZ model}\\\hline
N\'eel& $\D\geq1$& $e^{-t/\tone}$&$\tone\sim \D^{2}$\\
N\'eel& $\D<1$& $e^{-t/\ttwo}(\cos^2(\omega t)-const.)$&$\ttwo\sim \D^{-1}$\\
\hline
\hline
\multicolumn{3}{l}{Sec. \ref{sec:ll}: Luttinger model}\\\hline
 KG& $\D\gtrsim 0$& $e^{-t/\tone}$& $\tone=\frac{2}{K\pi\gap}$
\end{tabular}
\end{ruledtabular}
\footnotetext[1]{Valid in an \textit{intermediate} time regime (See  section \ref{sec:xxz}).}
\footnotetext[2]{Only rough correspondence, deviations of the order of $30\%$ are possible.}
\end{table*}

Here, it will be shown that in the long-time limit the antiferromagnetic order vanishes in all cases, hence, at least for this local quantity, thermalization is observed -- in a one-dimensional system no long-range order is possible at finite temperatures.  This does not necessarily imply thermalization for correlation functions. Indeed, in section \ref{sec:xxz} we present results which indicate the absence of thermalization in the spin-spin correlations (\ref{eq:spinspin}). However, the correlation functions exhibit somewhat slow relaxation dynamics and it is difficult to extract steady-state properties from the rather short accessible times that can be achieved numerically.

Nevertheless, interesting dynamical effects are present also at short times. Their
characterization as a function of the initial state and the interaction parameter will be investigated. The magnetic order parameter turns out to be a good observable for the quantitative extraction of non-trivial time scales. Here, where the initial state can be characterized by the gap parameter $\gap$ (more precisely, the inverse correlation length), one expects that the typical time scale of the relaxation dynamics is given by $\gap^{-1}$ and the length scales, which depend on the momentum distributions in the initial states, should be of the order of $u/\gap$, where $u$ is given by the velocity of quasi-particles (spin-waves).

In the solution of the quench dynamics for conformally invariant theories \cite{belavin-1984} of Calabrese and Cardy \cite{calabrese-2006,calabrese-2007} these qualitative arguments were put on a solid ground: The initial state enters into the framework of quantum field theory as a finite slab width, $\tau_e$, the extrapolation length which stands for the renormalization-group distance of the initial state from the fixed point of the gapped theory \cite{diel-1997}. To first order, this is given by the inverse gap, here $\tau_e\sim \gap^{-1}$. Using a conformal transformation, the slab geometry is mapped onto the semi-infinite plane, for which, by means of boundary conformal field theory \cite{cardy-1984}, the properties of the correlation functions can be extracted. 

The results of Calabrese and Cardy \cite{calabrese-2007} do apply to the quench in the XXZ model if the discussion is restricted to the low-energy modes in the gapless regime $|\D|\leq1$, here captured by the Luttinger model [see appendix \ref{sec:magnetismtheo}, Eq. (\ref{eq:llhamiltonian})]. For the staggered magnetization as a local observable the outcome is
\beq
\ms(t)\sim e^{-\frac{\pi K t}{2\tau_e}}\,,
\label{eq:conformallocal}
\eeq
where $\tau_e\sim\gap^{-1}$.

However, several remarks concerning the applicability of the conformal field theory results to the quench in the XXZ chain are in place. First, the initial state is treated on a perturbative level in terms of a renormalization-group distance from the fixed point and simply characterized by the gap parameter. It is questionable whether in this framework it is possible to correctly take into account the physics of the antiferromagnetic states, especially those close to the critical point (i.e. far from the antiferromagnetic fixed point), described by the sine-Gordon model. Second, within the field theory it is impossible to treat lattice effects, which are expected to emerge if the energy of the quasi-particles forming the initial state is of the order of the bandwidth $\L$ -- a situation which is realized for instance by the N\'eel state (\ref{eq:neel}). As a simple example of a lattice effect we presented the Rabi-oscillations in the two-spin system (\ref{eq:rabi}), with the frequency set by the magnetic exchange $J$. Macroscopic order parameter oscillations following a quantum quench  have been predicted to appear in a variety of systems  \cite{ polkovnikov-2002, sengupta-2004,hastings-2008,altman-2002,barankov-2004}. In this work we will characterize Rabi-like oscillations and investigate origins of dephasing in the presence of many-body correlations. A particular property of the quench in the XXZ chain illustrates the novel aspect of the non-equilibrium dynamics in many-body lattice models: The time-evolution of $\ms(t)$ is invariant under the change of sign $\Delta \rightarrow -\Delta$. Ferro- and antiferromagnetic Hamiltonians exhibit identical dynamics despite their completely different elementary excitations.
As a third point restricting the applicability of the conformal field theory result, we mention that a conformal theory does not capture the case of a parameter quench into the gapped phase, $\D>1$. Here this regime will be addressed using a sine-Gordon description of the XXZ model.

\subsection{Summary of the results}

\label{sec:summary}
The further content of the paper is organized as follows:  The non-equilibrium dynamics in the XX limit of the Heisenberg model, which can be solved in a simple way by means of Jordan-Wigner transformation, is analyzed in section \ref{sec:xx}.
Numerical results for the general case are given in section \ref{sec:xxz} and approximative approaches in sections \ref{sec:meanfield}-\ref{sec:ll}. In  section \ref{sec:spectral}  an exact diagonalization analysis of the spectrum of the XXZ model is carried out before presenting the conclusions.
In appendix \ref{sec:magnetismexp} we describe the experimental realization of quantum magnetic systems in optical lattices. The well-established properties of antiferromagnetic states and equilibrium phase transitions in the context of the Heisenberg model in one dimension are reviewed in appendix \ref{sec:magnetismtheo}. The description and an error analysis of the matrix product algorithm is provided in appendix \ref{sec:mps}.

Our results for the non-equilibrium dynamics of the staggered magnetization are summarized in Table \ref{tab:results}. We  find essentially  two types of relaxation dynamics: non-oscillatory dynamics, characterized by a relaxation time $\tone$, and oscillatory dynamics with  a frequency $\w$  and an associated relaxation time $\ttwo$. An important result is that for non-zero $\Delta$ we find a fundamentally new mode of many-body dynamics which always leads to {\em exponential} decay of the staggered moment regardless of whether the short-time dynamics is oscillatory or not. In contrast with the oscillation frequency, which is set by the exchange interaction, the relaxation time is an emergent scale generated by the highly correlated dynamics and hence cannot be simply related to the microscopic parameters. We find divergent relaxation times, $\tone\ra\infty$ in the limit $\D\to 0$ and $\ttwo\ra\infty$ if $\D\to\infty$. For the particular case of the N\'eel state, we find that the relaxation times essentially vanish in the vicinity of the critical point, $\D\gtrsim1$.

Table \ref{tab:results} also shows to what extent approximative methods, which take into consideration only a particular aspect of the Hamiltonian, are applicable to the non-equilibrium problem. The mean-field approximation for example leads to contradictions with the unbiased numerical results -- an algebraic decay for $\D\leq1$ and a non-vanishing asymptotic value of the staggered moment for $\D>1$ \cite{hastings-2008}. In the case of the initial N\'eel state, comparing the low-energy result of conformal field theory with the numerics, the immediate relaxation $\tone\approx0$ is in principle in agreement with $\gap\ra\infty$ in Eq. (\ref{eq:conformallocal}). However, the oscillations dominate the long-time dynamics, and are, as expounded before, not captured by the field theory. If the initial state is close to the critical point, an exponential relaxation similar to Eq. (\ref{eq:conformallocal}) is found, however, an additional algebraic prefactor appears to be present. 
In our treatment of the Luttinger model this effect is also not seen, but the results from conformal field theory (\ref{eq:conformallocal}) are reproduced.

\section{XX model, $\D=0$}
\label{sec:xx}
It is particularly illustrative to study the exactly solvable case of zero anisotropy ($\Delta=0$), where the Heisenberg Hamiltonian  (\ref{eq:xxzhamiltonianoned}) can be represented in terms of free spinless fermions with a cosine dispersion relation (\ref{eq:xxhamiltonian}). For free fermions the non-equilibrium dynamics can be solved analytically \cite{antal-1999}. We study two cases: first, the N\'eel state as the initial condition, second, the case of the initial spin-density-wave state.

\subsection{Initial N\'eel state, $\Delta_0=\infty$}
\begin{figure}[ht]
\centering
\includegraphics[width=\figwa\textwidth,angle=0]{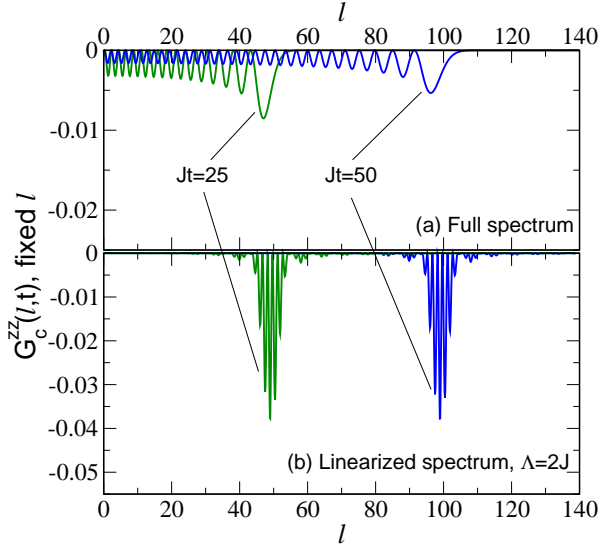}
\caption{\label{fig:szsz_xx}Correlation functions in the XX limit. Comparison of results obtained for full (a) and linearized (b) spectra. For the linearized spectrum we set $\L=2J$ as the effective bandwidth.}
\end{figure}

In the fermionic picture, the  N\'eel state reads as a charge density wave, 
\beq
\ket{\psi_0}=\prod_{\frac{-\pi}{2}<k\leq\frac{\pi}{2}}(c^\dagger_k+c^\dagger_{k+\pi})\ket{0}\,.
\eeq
The fermionic operators are easily represented in the Heisenberg picture,
\beq
c_k(t)=e^{it\e_kc_k^\dagger c_k}c_ke^{-it\e_kc_k^\dagger c_k}=c_ke^{-it\e_k}\,.
\eeq
Hence, the dynamics of the XX chain prepared in the N\'eel state, in analogy with the two-site model (\ref{eq:rabi}), takes the form of Rabi oscillations between charge-density waves with different sublattice magnetizations,
\beq
|\psi(t)\rangle=\prod_{-\frac{\pi}{2}<k\leq\frac{\pi}{2}}(e^{i\e_kt}c_k^\dagger+e^{-i\e_kt}c_{k+\pi}^\dagger)\ket{0}\,.
\eeq
The relaxation of the staggered magnetization can be seen as a dephasing process, driven by inhomogeneous Rabi frequencies in $k$-space,
\beq
m_s(t)&=&\frac{1}{N}\sum_{-\frac{\pi}{2}<k\leq\frac{\pi}{2}} e^{i2\epsilon_kt}\langle\psi_0|c_k^\dagger c_{k+\pi}|\psi_0\rangle\notag\\&=&\frac{1}{N}\sum_{-\frac{\pi}{2}<k\leq\frac{\pi}{2}}\frac{1}{2}e^{i2\epsilon_kt}\,.
\eeq
In the thermodynamic limit,
\beq
\ms(t)=\frac{1}{\pi}\int_0^\frac{\pi}{2}dk\cos(2t\epsilon_k)=\frac{1}{2}J_0(2Jt)\,,
\label{eq:msxx}
\eeq 
where $J_0$ denotes the zeroth Bessel function of the first kind. The underlying frequency distribution (\ref{eq:energydist}) ranges over a band of width $4J$,
\beq
f_{\ms}(\e)= \theta(2J-\e)\theta(\e+2J)\frac{1}{\sqrt{4J^2-\e^2}}\,,
\label{eq:xxdist}
\eeq
$\theta(\e)$ being the Heaviside function. High-energy modes with a vanishing velocity at the band edge, $ |\epsilon_k|=J$,  dominate the long-time limit of (\ref{eq:msxx}) and give rise to the oscillations with a frequency set by the bandwidth,
\beq
m_s(t)\xrightarrow{Jt \gg 1} \sqrt{\frac{1}{4 \pi J t}} \cos(2 J t - \frac{\pi}{4})\,.
\label{eq:msxxlt}
\eeq
The exponent of the $t^{-\frac{1}{2}}$ decay is a consequence of the quadratic  dispersion at $k=0$.
In the XX limit it is also possible to express the correlation function, $G_c^{zz}(\ell,t)$, in terms of Bessel functions,
\beq
G_c^{zz}(\ell,t)&=&\frac{\delta_{\ell,0}-1}{4\pi}\left[\int_{-\pi/2}^{\pi/2}dk\cos(k\ell)\cos(2t\epsilon_k) \right]^2\notag\\
&=&\frac{1}{4} \left( \delta_{\ell,0}-J_\ell^2(2 J t)\right)\,.
\eeq
This results in slowly decaying, spatially oscillating correlations,
\beq
&&G_c^{zz}(\ell,t)\xrightarrow{\ell\ll Jt} -\frac{1}{2\pi J t} \cos^2(2Jt-\ell\pi/2-\pi/4)\,.
\label{eq:szsz}
\eeq
Fig. \ref{fig:szsz_xx} shows how the correlations evolve within the light cone $\ell\leq 2 t$. The magnitude of the wave-front decays as a power law in time. The negative sign reflects spinon characteristics \cite{giamarchi-book} of the propagating correlations.

Although it is possible to carry out the analysis of the XX model without any approximation, it is useful to investigate the result of restriction to a particular part of  the spectrum. This provides information on the  range of applicability of low-energy theories, which are candidates for treating  the more complicated case of interacting systems.

In the case of the linearized theory [appendix \ref{sec:magnetismtheo}, Eq. (\ref{eq:xxlinearized})], the dynamics of the magnetization is characterized by oscillations with a $1/t$ decay and cutoff-dependent period,
\beq
m_s(t)=\frac{1}{\L t} \sin(\L t)\,.
\label{mslinearneel}
\eeq
The cutoff $\L$ gives the correct periodic behavior if it is equal to  the bare bandwidth ($\L=2J$). The oscillatory behavior, a consequence of the presence of the lattice, is indeed not captured in the continuum limit $\L/J\rightarrow \infty$, where the oscillations disappear. The power-law decay appears in the linear approximation being independent of the cutoff, but the exponent is overestimated by a factor of two compared to the case of the full dispersion. The energy distribution corresponding to the magnetization (\ref{mslinearneel}) is simply flat,
\beq
f_{\ms}(\e)= \theta(2\L-\e)\theta(2\L+\e)\,.
\eeq

A seemingly (in the context of equilibrium theories) unconventional approach is the development of the modes in the vicinity of the band edges,
\beq
H_Q=\sum_{Jk^2\leq\L} -J(1-k^2)\left\{c_k^\dagger c_k-c_{k+\pi}^\dagger c_{k+\pi}\right\}\,. 
\eeq
In the present case of non-equilibrium dynamics, we find, however, that the corresponding energy distribution,
\beq
f_{\ms}(\e)&=& \theta(\e+2J)\theta(\L-2J-\e)\frac{1}{\sqrt{2J+\e}}\\
&+&\theta(2J-\e)\theta(\L-2J+\e)\frac{1}{\sqrt{2J-\e}}\,,
\eeq
provides the correct long-time limit if the cutoff is sufficiently large,
\beq
\ms(t)\mathop{\sim}_{Jt\gg\L^{-1}} \frac{1}{\sqrt{t}}\cos(2Jt-\frac{\pi}{4})\,.
\eeq

We  now clearly understand the mechanism behind the dephasing process in the free-fermion models: Rabi oscillations are present if there is a sharp step at the edge of the band. The dephasing of the oscillations is algebraic, $t^{-\a}$, $\a=1$ if the frequency distribution is homogeneous and $\a=\frac{1}{2}$ in the case of the quadratic dispersion at the band edge. For the long-time behavior it is sufficient to stick to the modes at the edge of the band, the low-frequency part is effective only  at short times $t\sim J^{-1}$. The reason for such behavior is best illustrated in the analysis of the correlation functions for the linear spectrum. The result, as shown in Fig. \ref{fig:szsz_xx}, is a single  coherent spinon mode traveling the light cone $|2t-\ell|=0$. For the staggered magnetization as a local observable this means that it relaxes as soon as the spinon mode moves over more than one lattice distance $2t>1$. In contrast to the case of the full dispersion, there are no oscillations within the light cone. We note that this immediate decay is in agreement with the result of conformal field theory (\ref{eq:conformallocal}), which predicts  zero relaxation time for the N\'eel state due to its vanishing correlation length (inverse gap).

\subsection{Initial spin-density wave}
\label{sec:SDW}

As an introduction to our discussion of quenches from correlated antiferromagnetic states (i.e. quenches with $1<\Delta_0 < \infty$), we consider the time evolution of weakly antiferromagnetic spin-density-wave states  under the XX Hamiltonian [See appendix \ref{sec:magnetismtheo}, Eqs. (\ref{eq:xxhamiltonian}) and (\ref{eq:sdw})]. This section will provide a benchmark for the numerical results in section \ref{sec:xxz} and also discusses the applicability of effective low-energy theories to this quench.

The time evolution of the staggered magnetization $m_s(t)$ in the XX model starting from a SDW state 
at $t=0$ ($|\psi_0 \rangle=\prod_{-\pi/2<k \leq \pi/2}( u_kc_k^\dagger +v_kc_{k+\pi}^\dagger)|0\rangle$) is determined by the coefficients $u_k$ and $v_k$,
\begin{eqnarray}
m_s(t) &=& \frac{1}{N} \sum_{k=-\pi}^{\pi} \langle \psi_0 | c_{k+\pi}^\dagger(t) c_k(t) | \psi_0 \rangle \notag \\
&=& \int_{-\pi}^{\pi} \frac{dk}{2 \pi} e^{-i 2 \epsilon_k t} u_k v_k \ ,
\label{stagmag_SDW}
\end{eqnarray}
where we have taken the thermodynamic limit in the last equation. With the coefficients obeying $
u_k v_k = \frac{\Delta_s}{2 \sqrt{\epsilon_k^2+\Delta_s^2}}$, the dephasing process in energy representation reads
\beq
m_s(t)=\frac{1}{\pi}\int_{-J}^0d\e \frac{\cos(2t\e)}{\sqrt{J^2-\e^2}}\frac{\Delta_s}{\sqrt{\epsilon^2+\Delta_s^2}}\,.
\label{eq:sdwfrequency}
\eeq
For a weak SDW state ($\Delta_s \ll 1$) there are two main contributions to the integral in Eq. (\ref{eq:sdwfrequency}). The first comes from the  Fermi points $\e=0$, whereas the second originates in the square root singularities  at $\e=\pm J$.  We write these two contributions separately,
\begin{eqnarray}
&&m_s(t) \approx\! \frac{\Delta_s}{\pi J} \, K_0(2 \Delta_s t)+\frac{\Delta_s}{2 J} \, J_0(2 J t) \label{eq:mssdw}\\
&&\xrightarrow{t\gg\gap^{-1}} \! \frac{1}{\sqrt{4 \pi J t}} \left\{ \sqrt{\frac{\Delta_s}{J}} e^{-2 \Delta_s t} + \frac{\Delta_s}{J} \cos(2Jt-\pi/4) \right\} \notag\,.
\end{eqnarray}
In comparision with the case of the initial N\'eel state, in addition to identical algebraically decaying oscillations (\ref{eq:msxx}) a non-oscillatory decay stemming from the low-energy part of the spectrum is obtained. This exponential behavior with an algebraic prefactor is characterized by the relaxation time $\tau=(2\gap)^{-1}$. Hence, for $t >\Delta_s^{-1} \ln(J/\Delta_s)$ the oscillations on top of the non-oscillatory decay dominate the order-parameter dynamics. Nevertheless, unlike the case of the initial N\'eel state, the low-energy modes contribute to the non-equilibrium dynamics over significant periods of time.

\section{Interaction quench in the XXZ-model -- numerical study}
\label{sec:xxz}
In this section we first study the quench in the XXZ model starting from the N\'eel state. Subsequently ground states of the XXZ models at finite $\D=\D_0$ will be considered.

Unlike for $\Delta = 0$, the problem is no longer analytically treatable and we have to resort to numerical techniques. In the iMPS algorithm (appendix \ref{sec:mps}) we use 2000 states and a second-order Suzuki-Trotter decomposition with a time step $\delta\sim 10^{-3} J^{-1}$ for large $\Delta$ and up to 7000 states with $\delta\sim 10^{-2}J^{-1}$  for small $\Delta$. An {\it intermediate} time regime $Jt\lesssim16$ can be reached, which exceeds in general greatly the short transient time.

\begin{figure}[ht]
\centering
\includegraphics[width=\figwa\textwidth,angle=0]{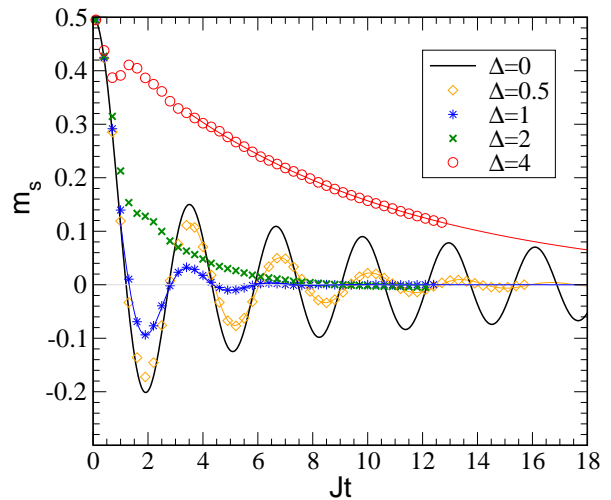}
\caption{\label{fig:ms_all}\figpreamble Dynamics of the staggered magnetization $\ms(t)$ in the XXZ chain initialized in a N\'eel state. Symbols  correspond to numerical results, lines represent analytical results or fits by corresponding laws (\ref{eq:damping}) and (\ref{eq:relaxation}) (see text). 
}
\end{figure}

\begin{figure}[ht]
\centering
\includegraphics[width=\figwa\textwidth,angle=0]{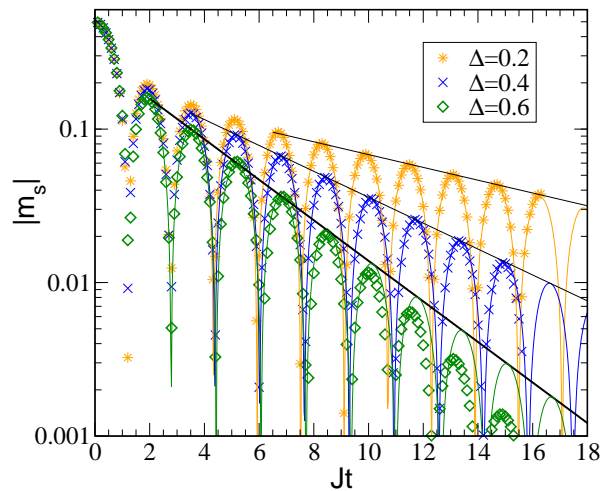}
\caption{\label{fig:ms_weak}\figpreamble Analysis of the decay of the oscillations in the XXZ model by plotting  the absolute value of the staggered magnetization on a logarithmic scale. Symbols represent numerical results, solid curves correspond to fits by the function (\ref{eq:damping}), straight lines point out the exponential decay. For anisotropies close to zero ($\Delta=0.2,0.4$) an exponential law is obeyed for large periods of time in the numerically accessible time window, while for $\Delta=0.6$ the simulation shows an acceleration of the decay after a few oscillations.}
\end{figure}

\begin{figure}[ht]
\centering
\includegraphics[width=\figwb\textwidth]{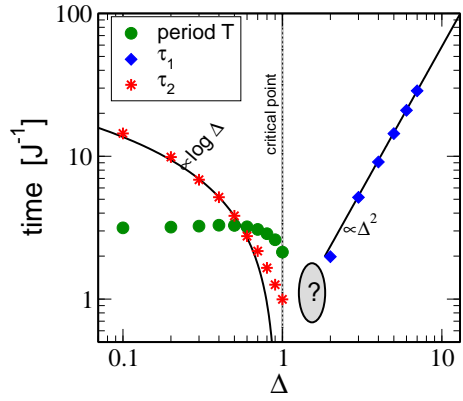}
\caption{\label{fig:relaxtime_neel}\figpreamble Relaxation times and oscillation period $T=\frac{2\pi}{\omega}$ as a function of anisotropy in the XXZ model for the N\'eel initial state. Logarithmic or algebraic laws are emphasized  by solid lines. In the region close to the critical point of the XXZ model (indicated by the question mark) it becomes impossible to extract a relaxation time from the numerical results.}
\end{figure}

\subsection{Initial N\'eel state, $\Delta_0=\infty$}

\begin{figure}[ht]
\centering
\includegraphics[width=\figwa\textwidth,angle=0]{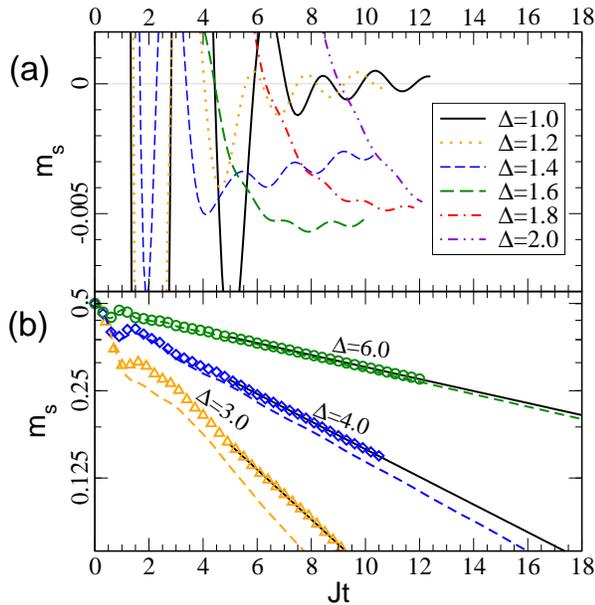}
\caption{\label{fig:ms_strong}\figpreamble (a) Focus on the XXZ chain close to the critical point $\Delta=1$, where a steep decay of the initial magnetization is followed by a rather slow relaxation on tiny magnitudes which does not fit  either of the generic behaviors (\ref{eq:damping},\ref{eq:relaxation}).
(b) Comparison of the XXZ chain (symbols) and the XZ chain (dashed lines) for strong anisotropies, solid lines correspond to an exponential fit. The dynamics of the staggered magnetization of the XXZ and XZ chains converge towards each other in the large-$\Delta$ limit.
}
\end{figure}

\begin{figure}[t]
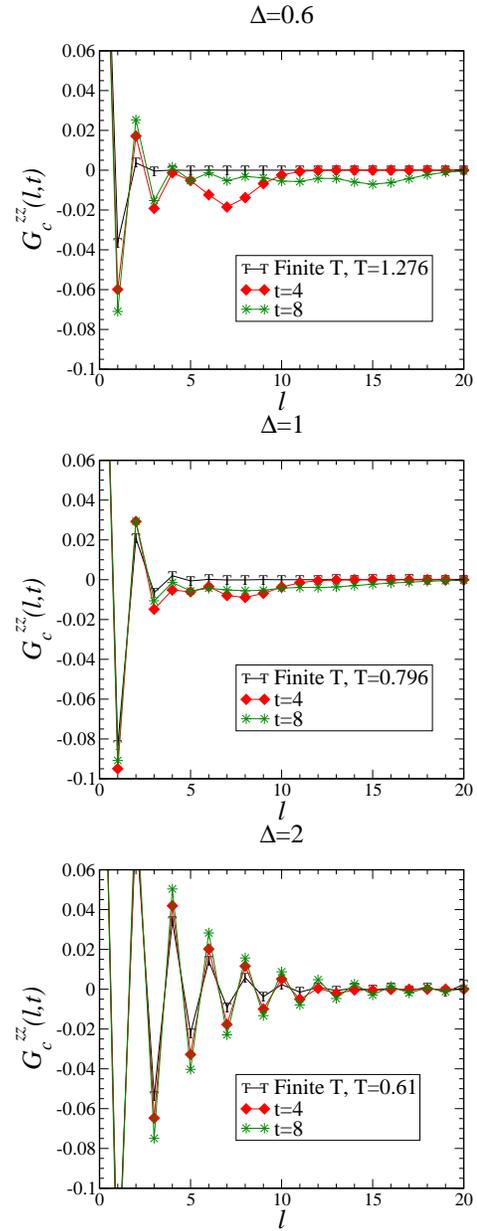

\centering
\includegraphics[width=\figwb\textwidth,angle=0]{szsz_neel_d06.eps}
\includegraphics[width=\figwb\textwidth,angle=0]{szsz_neel_d1.eps}
\includegraphics[width=\figwb\textwidth,angle=0]{szsz_neel_d2.eps}
\caption{
\label{fig:corr} The correlation functions obtained using iMPS for the initial N\'eel state. Symbols $T$ denote quantum Monte Carlo results for the XXZ model at equilibrium at a temperature fixed by the energy of the non-equilibrium system. 
}
\end{figure}

An overview of the results for the initial N\'eel state is presented in Fig.~\ref{fig:ms_all}. For small anisotropies we find oscillations of the order parameter similar to those in the XX limit, but with the decay time decreasing upon approaching the isotropic point $\Delta=1$. In the easy-axis regime $\Delta>1$ of the XXZ model, the relaxation slows down again for increasing $\Delta$ and we observe non-oscillatory behavior for $\Delta\gg1$.

Fig. \ref{fig:ms_weak} focuses on easy-plane anisotropy $0<\Delta<1$.
The results for $0<\Delta\leq0.4$ are well described, for accessible time scales, by exponentially decaying oscillations
\beq
\ms(t)\propto e^{-t/\ttwo}\cos(\w t +\phi)\,.
\label{eq:damping}
\eeq
The oscillation frequency is almost independent of the anisotropy, while the relaxation time $\ttwo$ increases with decreasing $\Delta$.  Logarithmic divergence of the relaxation time in the limit $\D\to 0$ is suggested by the fit shown in Fig. \ref{fig:relaxtime_neel}. The picture is less clear closer to the isotropic point. For the range $0.5\leq\Delta<1$ there appears to be an additional time scale after which the oscillations start to decay even faster than exponentially, simultaneously the period of the oscillations is reduced. Therefore, the relaxation times plotted in Fig. \ref{fig:relaxtime_neel} are only valid within an intermediate time window, whose width shrinks upon approaching the critical point.

For intermediate easy-axis anisotropies $1\leq\Delta\leq3$, the magnetization does not reach a stable regime within the numerically accessible time window [Fig. \ref{fig:ms_strong}(a)]. The complicated behavior of $m_s(t)$ in this parameter range can be ascribed to the interplay of processes at all energy scales. Nevertheless, the numerical data suggest that the relaxation is fastest close to the isotropic point, in the range between $\Delta=1$ and $\Delta=1.6$. A simple generic type of behavior is recovered for large anisotropies $\Delta\gtrsim3$. The numerical data in Fig. \ref{fig:ms_strong}(b) indicates exponential relaxation of the staggered magnetization
\beq
\ms(t)\propto e^{-t/\tone}\,.
\label{eq:relaxation}
\eeq
The relaxation time scales roughly quadratically with $\Delta$ (Fig. \ref{fig:relaxtime_neel}). Oscillations do persist on top of the exponential decay, but they fade out quickly.

We briefly describe the relaxation of the spin-spin correlation functions (\ref{eq:spinspin}) as presented in Fig. \ref{fig:corr}. A more detailed study of these  has been carried out by Manmana \etal ~\cite{manmana-2009}. For weak interactions (e.g. $\Delta=0.6$) the dynamics of correlation functions is still dominated by the spinon mode  moving according to the light-cone \cite{lieb-1972,calabrese-2005} set by the spin-wave velocity $u$ [See appendix \ref{sec:magnetismtheo}, Eq. (\ref{eq:llparameters})], as it is the case at $\D=0$ (Eq. (\ref{eq:szsz}), Fig. \ref{fig:szsz_xx}). For larger $\D$, this mode is smeared off, instead, antiferromagnetic correlations build up. The strength of the short-range antiferromagnetic correlations increases as the anisotropy $\Delta$ is augmented. With the numerical method, however, we are unable to reach sufficiently long times to calculate the quasi-stationary correlation length. It becomes nevertheless clear that the correlations cannot be described in terms of a thermal ensemble. We evaluated the equilibrium correlation functions at a temperature corresponding to the energy of the system by means of quantum Monte Carlo simulations \footnote{We have been using the ALPS code \cite{alps-2007} for the directed loop algorithm in the stochastic series expansion representation \cite{sandwik-1991,alet-2005}.}. The resulting correlation functions depicted in Fig. \ref{fig:corr} decay considerably faster than the non-equilibrium ones.

\subsection{Initial antiferromagnet, $1<\Delta_0<\infty$}
\begin{figure}[b]
\centering
\includegraphics[width=\figwb\textwidth,angle=0]{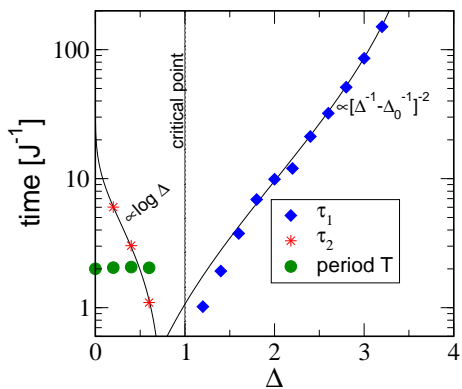}
\caption{\label{fig:relaxtime_xxz4} \figpreamble Relaxation times and oscillation period $T=\frac{2\pi}{\omega}$ as a function of anisotropy in the XXZ model for the system prepared in the ground state for $\D_0=4$. Logarithmic or algebraic laws are emphasized  by solid lines.}
\end{figure}

The N\'eel state is an entirely classical state with no quantum correlations. In order to generalize our results, we first study the case of small but finite correlations starting from the ground state for $\D_0=4.0$. We find that the picture gained from the initial N\'eel state remains qualitatively valid -- the dynamics of $\ms(t)$ is very similar to that in the case of the initial N\'eel state (Fig. \ref{fig:ms_all}). The corresponding relaxation times and periods are plotted in Fig. \ref{fig:relaxtime_xxz4}.  For $\D$ close to zero, the behavior of $\ttwo$ is again close to a logarithmic law and the divergence of the relaxation time for $\D \rightarrow\D_0$ goes like  $\tone\propto |\frac{1}{\Delta}-\frac{1}{\D_0}|^{-2}$. 

\begin{figure}[bh]
\centering
\includegraphics[width=\figwa\textwidth,angle=0]{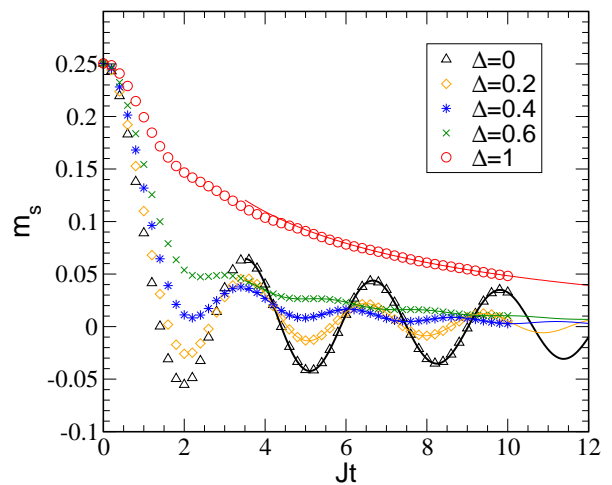}
\caption{\label{fig:ms_15} Dynamics of the staggered magnetization $\ms(t)$ in the XXZ chain prepared in an antiferromagnetic ground-state of the XXZ Hamiltonian with $\Delta=\Delta_0=1.5$. Symbols  correspond to numerical results, lines represent analytical results or fits by corresponding laws (\ref{eq:mstweak}), (\ref{eq:af2}) and (\ref{eq:relaxation}) (see text).}
\end{figure}
\begin{figure}[t]
\centering
\includegraphics[width=\figwb\textwidth,angle=0]{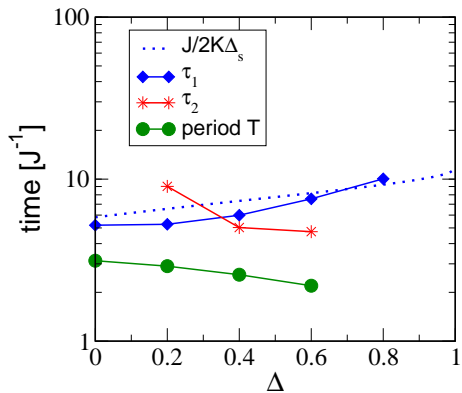}
\caption{\label{fig:relaxtime_weak}Relaxation times and oscillation period $T=\frac{2\pi}{\omega}$ as a function of anisotropy in the XXZ model for the system prepared in the ground state at $\D_0=1.5$. Solid lines are guides to the eye. $\tone$ is comparable to $(J/K\gap)$.}
\end{figure}

We expect a qualitatively different behavior for a weakly ordered (more strongly correlated) initial state. In section \ref{sec:xx} we have seen that for an initial spin-density-wave state and the XX Hamiltonian, in  addition to the algebraically decaying oscillations, an exponential relaxation exists, whose relaxation rate is proportional to the gap of the initial state. In Fig. \ref{fig:ms_15}, where we show the results for the quench from an initial state with $\D_0=1.5$, oscillations are found on top of non-oscillatory relaxation. At $\D=0$, for sufficiently large $t$, the dynamics is similar to the SDW result (\ref{eq:mssdw}),
\beq
\ms(t) \sim C_1  t^{-\frac{1}{2}}e^{-t/\tau_{\noss}} +C_2 t^{-\frac{1}{2}}\cos(\omega t +\phi)\,,
\label{eq:mstweak}
\eeq
to very high accuracy, despite the fact that the spin-density wave is a different wave function than the ground state of the XXZ chain. The relaxation time $\tau_{\noss}\approx 5.1J$ is slightly smaller than the one predicted by the SDW calculations, $(2\Delta_s)^{-1}\approx 5.8J$. The difference may be explained by the importance of short-range effects which are supposed to contribute to the non-equilibrium dynamics. As illustrated in Fig. \ref{fig:groundstate}, the correlations decay much faster at shorter distances than in the large distance asymptotics. 

For $\D\gtrsim0$, in correspondence with the result for the initial N\'eel state, we find that the oscillations are exponentially damped, while the non-oscillatory part remains qualitatively the same as in the XX limit,
\beq
\ms(t) \sim C_1  t^{-\frac{1}{2}}e^{-t/\tau_{\noss}} +C_2 e^{-t/\tau_{\oss}}\cos(\omega t +\phi)\,.
\label{eq:af2}
\eeq
In Fig. \ref{fig:relaxtime_weak} we plot the fitting parameters for small  $\Delta$ ($0<\D\leq0.6$) where formula (\ref{eq:af2}) is well obeyed. $\tone$ behaves similarly to $(J/K\gap)$ -- a law which is the natural extension of the non-interacting SDW result (\ref{eq:mssdw}) to finite anisotropies using the same scaling as derived for Luttinger model [see section \ref{sec:ll}, Eq. (\ref{LLresult})]. The logarithmic behavior of $\ttwo$, apparent for $\D_0\gg1$, is not observed here.  Oscillatory and non-oscillatory terms are superimposed. In the non-oscillatory  term of (\ref{eq:af2}) absence of the algebraic prefactor, as suggested by the field theoretical-result (\ref{eq:conformallocal}), can be clearly excluded on the basis of the numerical results. Pure exponential law (\ref{eq:relaxation}) is however found for $\D\gtrsim1$. The intermediate regime $0.6\lesssim\D\lesssim 1$ can not be described by either of the laws (\ref{eq:relaxation}),(\ref{eq:af2}).

\section{Mean field}
\label{sec:meanfield}
Time-dependent mean field theory is one possibility to treat the dynamics of the XXZ model approximately. The mean-field approximation of the Hamiltonian (\ref{eq:xxzcharge}) at an instant of time $t$, is defined by expanding the interaction term to linear order in fluctuations $\delta n_j$ around the mean density, $n_j =\langle n_j \rangle +\delta n_j$, and by setting $\langle n_j \rangle = 1/2+ (-1)^j\ms$,
\begin{equation}
H_{MF}(t)=-J \sum_{k=-\pi}^\pi \left( \cos(k) c^\dagger_k c_k + 2 \Delta \ms(t) c^\dagger_{k+\pi} c_k \right) \,,
\label{eq:mf}
\end{equation}
where the mean-field staggered magnetization $\ms(t)$ has to be determined self-consistently. For developing an intuition it is worthwhile to imagine the dynamics of pseudo-spins in k-space by defining pseudo-spin operators $\sigma_k^z=c^\dagger_k c_k-c^\dagger_{k+\pi} c_{k+\pi}$ and $\sigma_k^x=c^\dagger_{k+\pi} c_k+c^\dagger_k c_{k+\pi}$. Note that these momentum-space pseudo-spins are different from the original spins on the chain. In pseudo-spin representation the staggered magnetization is given by the average $x$-projection per pseudo-spin, $\ms=\frac{1}{N} \sum_{k=-\pi/2}^{\pi/2} \langle \sigma_k^x \rangle$, and the mean-field Hamiltonian can be written as
\begin{equation}
H_{MF}(t) = -J \sum_{k=-\pi/2}^{\pi/2} \big( \cos(k) \, \sigma_k^z + 2 \Delta \, \ms(t) \, \sigma_k^x \big)\,,
\label{MF_hamiltonian_pseudospin}
\end{equation}
The N\'eel state as an initial condition corresponds to all pseudo-spins pointing in $x$-direction at $t=0$. Then they start to precess due to a Zeeman field that depends on the instantaneous average orientation of the $x$-projection of the spins. In these terms it is easy to understand the evolution of the staggered magnetization $\ms(t)$ for $\Delta=0$. We simply have a collection of independent pseudo-spins subject to constant Zeeman fields $J\cos k$ in the $z$-direction. Because the field magnitude varies from spin to spin over a bandwidth, they precess at frequency  $2 J$. Since the band of precession frequencies is continuous, the spins gradually dephase leading to the $1/\sqrt{t}$ decay (\ref{eq:msxxlt}) of the oscillation envelope of $\ms(t)$.

\begin{figure}[t]
\centering
\includegraphics[width=\figwa\textwidth,angle=0]{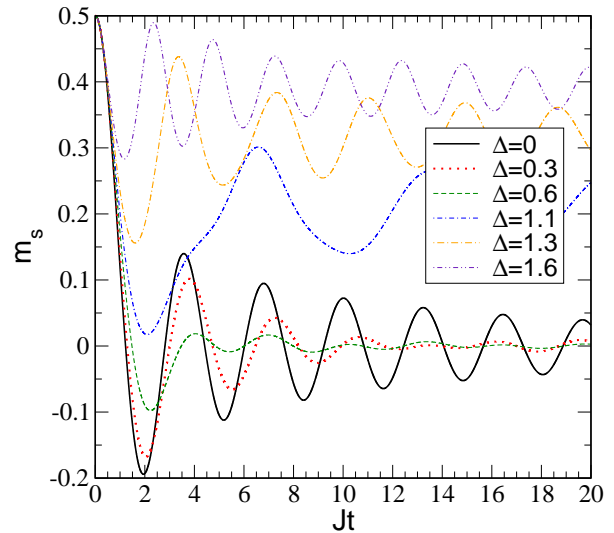}
\caption{\label{fig:mf}Numerical solution of the mean field equations for the Hamiltonian (\ref{MF_hamiltonian_pseudospin}) with the initial condition $\ms(0)=\frac{1}{2}$.}
\end{figure}

The situation is more complicated in the case of $\Delta \neq 0$. Now there is also a field in the $x$-direction, which is the same for all spins but changes in time according to the instantaneous orientation of the spins. To lowest order in $\Delta$, i.e. setting $\ms(t)=m_0(t)=J_0(2 J t)/2$ in the Hamiltonian (\ref{MF_hamiltonian_pseudospin}), the additional Zeeman field in $x$-direction tilts the precession axis, giving rise to a smaller average $x$-projection of the spins and thus leading to a faster decay of $m(t)$. The numerical results for the time evolution of the staggered magnetization according to (\ref{MF_hamiltonian_pseudospin}) are shown in Fig. \ref{fig:mf} for different values of $\Delta$. Finite $\D$ leads to accelerated dephasing of the oscillations very much like in the unbiased calculations (Sec. \ref{sec:xxz}). However, the asymptotic law as extracted from the numerical solution by Hastings and Levitov \cite{hastings-2008} for $0<|\D|\leq1$  exhibits algebraically decaying oscillations with a $t^{-\frac{2}{3}}$ envelope,
\beq
\ms(t)\sim t^{-\frac{2}{3}}\left\{C_1 \cos(\w_1t+ \phi_1) + C_2 \cos(\w_2t+ \phi_2)\right\}\,,
\eeq
$\w_1=2J$, $\w_2=\sqrt{1-\D^2}$. This algebraic decay, as well as the two frequencies, which lead to a revival phenomenon \cite{hastings-2008}, is in contradiction with the MPS calculations for the full Hamiltonian (\ref{eq:xxzhamiltonianoned}). For $\Delta>1$, the staggered magnetization saturates to a nonzero value for $t \rightarrow \infty$, which is presumably also an artifact of the mean-field approach not corroborated in the unbiased treatment. We conclude that the approach provides only a very rough picture of the order-parameter dynamics, which confirms the importance of collective effects,  apparently not captured by the effective non-interacting mean-field Hamiltonian (\ref{eq:mf}).

\section{XZ-model -- effective description for $\D\gg1$}
\label{sec:xz}

In this section we study the time evolution of the staggered magnetization $m_s(t)$ following a quench from the N\'eel state in the analytically treatable XZ model. This serves as a complementary analytical approach to the numerical investigation of the quench dynamics in the XXZ model in the regime of large anisotropies $\Delta \gg1$ and allows of a discussion of the long-time asymptotic behavior of $m_s(t)$. The XZ model is defined by the Hamiltonian
\begin{eqnarray}
H_{\text{XZ}} &=& J \sum_j \left\{ 2 S_j^x S_{j+1}^x + \Delta  S_j^z S_{j+1}^z \right\} \label{XZhamiltonian} \\
&=& H_{\text{XXZ}}+ \frac{J}{2} \sum_j \left\{ S_j^+ S_{j+1}^+ + S_j^- S_{j+1}^-\right\} \notag\, .
\end{eqnarray}
At equilibrium the XZ model exhibits a quantum phase transition at $\D=\Delta_c=2$ which separates two gapped phases with antiferromagnetically ordered ground states in z-direction for $\Delta>\Delta_c$ and in x-direction for $\Delta<\Delta_c$. It differs from the XXZ model (\ref{eq:xxzhamiltonianoned}) by terms violating the conservation of $S^z_\text{tot}=\sum_j S^z_j$, but has the advantage of being analytically diagonalizable. In the following we will prove that the staggered magnetization in this model vanishes for all finite $\Delta>\Delta_c$ in the long-time limit after a quench from the N\'eel state and calculate the exact time evolution of $m_s(t)$ semi-analytically up to times $J t \approx 100$, thus going beyond the time window accessible by the MPS calculation for the XXZ chain.  

\begin{figure}[b]
\centering
\includegraphics[width=\figwa\textwidth,angle=0]{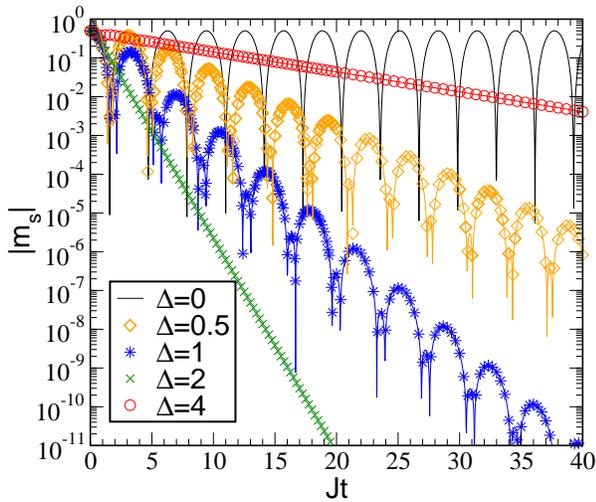}
\caption{\label{fig:ms_xz_all}\figpreamble Dynamics of the staggered magnetization $\ms(t)$ in the XZ chain initialized in a N\'eel state. Symbols  correspond to numerical results, lines represent analytical results or fits by corresponding laws (see text).
}
\end{figure}

\begin{figure}[b]
\centering
\includegraphics[width=\figwb\textwidth]{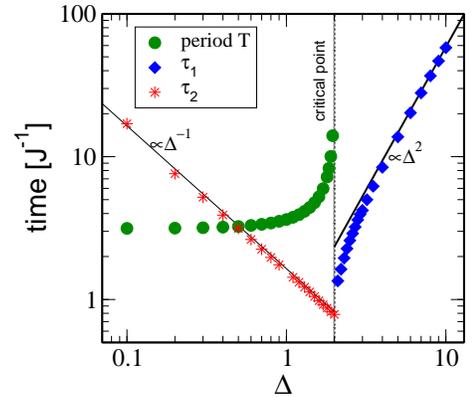}
\caption{\label{fig:relaxtime_xz}\figpreamble Relaxation times and oscillation period $T=\frac{2\pi}{\omega}$ as a function of anisotropy in the XZ model. Algebraic laws are emphasized by solid lines.}
\end{figure}

Using the Jordan-Wigner transformation for $S_j^x$ and $S_j^z$ and going over to momentum representation, the Hamiltonian of the XZ model (\ref{XZhamiltonian}) takes the form
\begin{eqnarray}
H_\text{XZ} &=& \frac{J}{2} \sum_{k=-\pi}^{\pi} \left\{ (\Delta+2) \cos(k) a_k^\dagger a_k \right. \notag \\
 &+& \left. \frac{i}{2} (\Delta-2) \sin(k) \left(a_k^\dagger a_{-k}^\dagger+a_k a_{-k} \right) \right\},
\label{XZ_JW}
\end{eqnarray}
with $a_k^\dagger$ and $a_k$ denoting respectively creation and annihilation operators of spinless Jordan-Wigner fermions with quasi-momentum k. This Hamiltonian can be diagonalized by the Bogoliubov transformation,
\beq
\begin{bmatrix} a_{-k} \\ a_k^\dagger \end{bmatrix} =&& \begin{bmatrix} \cos \theta_k & -i \sin \theta_k \\ -i \sin \theta_k & \cos \theta_k \end{bmatrix}  \begin{bmatrix} \alpha_{-k} \\ \alpha_k^\dagger \end{bmatrix} = M_\Delta \begin{bmatrix} \alpha_{-k} \\ \alpha_k^\dagger \end{bmatrix},\notag\\
\tan 2\theta_k &=& \frac{2-\Delta}{2+\Delta} \tan k\,,
\eeq

which, maps (\ref{XZ_JW}) to a model of free fermions,
\begin{equation}
H= \sum_k \varepsilon_k \left(\alpha_k^\dagger \alpha_k-\frac{1}{2}\right)
\end{equation}
with a dispersion $\varepsilon_k=J \sqrt{1+\Delta^2/4+\Delta \cos 2k}$. 

Since the initial N\'eel state is the ground state of (\ref{XZ_JW}) with $\Delta=\Delta_0 \rightarrow \infty$, it is convenient to express the time-dependent (Heisenberg) Jordan-Wigner fermion operators $a_k(t)$ in terms of the Bogoliubov quasiparticle operators $\alpha_k^0$ which diagonalize the Hamiltonian (\ref{XZ_JW}) for the initial value $\Delta=\Delta_0$,
\begin{equation}
\begin{bmatrix} a_{-k}(t) \\ a_k^\dagger(t) \end{bmatrix}  = M_\Delta \begin{bmatrix} e^{-i \varepsilon_k t} & 0 \\ 0 & e^{i \varepsilon_k t} \end{bmatrix} M^{-1}_\Delta M_{\Delta_0} \begin{bmatrix} \alpha^0_{-k} \\ {\alpha^0_k}^\dagger \end{bmatrix}. 
\label{eq:bogtrafo}
\end{equation}
This reduces the computation of correlation functions at arbitrary time to the evaluation of ground-state expectation values.

To calculate the time evolution of the staggered magnetization $m_s(t)$ following a quench in the XZ model, we define the two-spin correlation function,
\beq
C(\ell,t)=(-1)^\ell \langle \psi_0 | S^z_0(t) S^z_\ell(t) | \psi_0 \rangle \ ,
\eeq
from which the square of the staggered magnetization is obtained by taking the infinite-range limit,
\beq
m_s^2(t) = \lim_{\ell\rightarrow \infty} C(\ell,t) \ .
\label{eq:ms2}
\eeq
In the fermionized picture of the XZ model the two-spin correlator takes the form
\begin{equation}
\langle S_0^z S_\ell^z \rangle = \frac{1}{4} (-1)^\ell \langle A_0 B_1 A_1\dots B_{\ell-1} A_{\ell-1} B_\ell \rangle \ ,
\label{correl}
\end{equation}
with  $A_j=a^\dagger_j+a_j$ and $B_j=a^\dagger_j-a_j$ being the Majorana operators at lattice site $j$  \cite{lieb-1961}. Using Wick's theorem, this correlation function can be expressed as a Pfaffian of pairwise contractions \cite{barouch-1971b}. For the quench problem studied here, the explicit form of these contractions follows from \eqref{eq:bogtrafo} and is given by
\begin{eqnarray}
\langle A_j A_i \rangle &=&\langle B_j B_i \rangle\\\notag
&=& \int_{-\pi}^\pi \frac{dk}{2\pi} \; e^{-i k (j-i)} \sin 2 \varepsilon_k t \, \sin2\phi_k  \label{AA}\,,\mbox{for $i\neq j$,}\\
\langle A_j B_i \rangle &=& \int_{-\pi}^\pi \frac{dk}{2\pi} \; e^{-i k (j-i)} e^{i 2 \theta_k} ( \cos 2\phi_k  
 \notag\\  &&- i \sin 2 \phi_k \, \cos 2 \varepsilon_k t) \ , \label{AB} 
\end{eqnarray}
with $\phi_k=\theta_k-\theta_k^0$ (see also \cite{sengupta-2004}, where identical expressions have been derived for the transverse-field Ising model). We have taken the thermodynamic limit and converted the sums into integrals in the expresions above.
In the limit $t\rightarrow\infty$ for $\Delta>\Delta_c$ the evaluation of \eqref{eq:ms2} reduces to the computation of a Toeplitz determinant, since the contractions \eqref{AA} of the $A_j$'s and $B_j$'s among themselves vanish. Szeg\"o's theorem can then be used to calculate the asymptotics of the Toeplitz determinant, yielding the result
\begin{equation}
\lim_{t\rightarrow\infty} C(\ell,t) \ \mathop{\approx}^{\ell\gg1} \ \frac{1}{4} \! \left( \frac{1+\sqrt{1-4/\Delta^2}}{2}\right)^\ell\,.
\end{equation}
Thus, after a quench from the N\'eel state in the XZ model, the staggered magnetization vanishes for all finite $\Delta>\Delta_c$ at large times.

At finite times, when the contractions (\ref{AA}) do not vanish, the Pfaffian representing the two-spin correlator (\ref{correl}) can be evaluated numerically at arbitrary times for a given distance. Due to the so called light-cone effect \cite{lieb-1972,calabrese-2007}, two spins at a distance $\ell$ are not causally connected at times smaller than $u t  < \ell/2$, since the correlation length of the initial N\'eel state is zero. Here $u$ denotes the maximum (classical) speed of quasiparticles, which in the XZ model is given by $u = \max_k (\partial_k \varepsilon_k)=2 J$. Exploiting this light-cone effect, the staggered magnetization can be calculated in terms of a finite-range correlation function,
\begin{equation}
m_s^2(t) \Big|_{2 J t < \frac{\ell}{2}} \approx C(\ell,t) \ .
\end{equation}
This method significiantly reduces the computational effort at short times. We remark however that the light cone is not completely sharp in quantum-mechanical systems \cite{calabrese-2007}. Nevertheless, for practical finite-precision calculations the infinite-range limit of the two-spin correlator is reached for distances just a few lattice sites beyond the light cone.

The results for the time evolution of the staggered magnetization following a quench from the N\'eel state in the XZ model are displayed in Fig. \ref{fig:ms_xz_all}.
As is the case for the XXZ chain, an explicit analytical expression for $m_s(t)$ in the XZ model can be derived for a quench to $\Delta=0$, which is given by $m_s(t)=0.5 \cos^2(J t)$.
For $\Delta<\Delta_c$, the numerical data for $m_s(t)$ at large times fits very well exponentially decaying oscillations of the form
\beq
m_s(t)\propto e^{-t/\ttwo}(\cos^2(\omega t)-const.)
\eeq
In this regime, the behavior of $m_s(t)$ in the XZ model is qualitatively different from that in the XXZ model, as can be seen from the period of the magnetization oscillations. In the XZ model the period diverges at the critical point (see Fig. \ref{fig:relaxtime_xz}) , whereas it becomes smaller upon approaching the isotropic point in the XXZ model (see Fig. \ref{fig:relaxtime_neel}). Furthermore, the critical point exactly marks the crossover between oscillatory and non-oscillatory behavior of $m_s(t)$ in the XZ model.

For $\Delta \geq \Delta_c$, the staggered magnetization decays exponentially in the XZ model and shows no oscillations at large times. Interestingly, the numerical results for $m_s(t)$ in the XXZ  and XZ models are almost indistinguishable at large anisotropies $\Delta \gg 1$, as can be seen from Fig. \ref{fig:ms_xz_all}. 
We have extracted the relaxation times from exponential fits to the numerical data, obtaining a clearly pronounced minimum right at the isotropic point (see Fig. \ref{fig:relaxtime_xz}). The relaxation time scales as $\ttwo\propto\Delta^{-1}$ for $\Delta \leq \Delta_c$ and as $\tone \propto\Delta^2$ for $\Delta\gg\Delta_c$.

\section{Gapless theory -- Luttinger model}
\label{sec:ll}

In the analysis of the XX limit (Section \ref{sec:xx}) it became evident that, if the initial gap is sufficiently small, the non-oscillatory relaxation of the order-parameter dynamics is determined by low-energy modes, which motivates the application of the Luttinger model to a quench to the gapless phase $\Delta<1$ of the XXZ model.

In Section \ref{sec:magnetismexp} the Luttinger model, $H_{LL}= \frac{u}{2 \pi} \int dx \left\{ K \left( \pi \Pi(x) \right)^2 + \frac{1}{K} \left( \partial_x \phi(x) \right)^2 \right\}$, has been introduced as a low-energy effective theory for the XXZ chain in the easy-plane regime. The bosonized form of the staggered magnetization is given by $m_s \sim \langle \cos(2 \phi) \rangle_{x=0}$, where made use of the translational invariance. The remaining problem amounts to computing the time evolution of $\langle \cos(2 \phi) \rangle$, starting from a state where the field $\phi$ is initially pinned at $0$ or $\pi/2$. We remark that this problem is essentially the dual of the dephasing problem studied in \cite{bistritzer-2007}, and thus we expect an exponential decay of $m_s$ with a characteristic time scale $\tau \sim J/(K \Delta_s)$. A convenient technique for solving this problem is the truncated Wigner method \cite{polkovnikov-2003}, which is exact for quadratic Hamiltonians such as (\ref{eq:llhamiltonian}). Using this approach, the time-dependent expectation value of the staggered magnetization can be written as a functional integral over the Wigner transform $\varrho_W(\phi_0,\dot{\phi}_0)$ of the initial density matrix:
\begin{eqnarray}
\langle \cos(2 \phi) \rangle &=& \int \mathcal{D} \phi(t) \int \mathcal{D}(\phi_0,\dot{\phi_0}) \, \varrho_W(\phi_0,\dot{\phi}_0) \, \notag\\ &&\times\cos(2 \phi) \, \delta(\ddot{\phi}-u^2 \partial_x^2 \phi)  \\
&=& \int \mathcal{D}(\phi_0,\dot{\phi}_0) \, \varrho_W(\phi_0,\dot{\phi}_0) \cos(2 \phi_\text{cl}(x,t))\notag
\end{eqnarray}
Here, the functional $\delta$-distribution ensures that one integrates only over solutions of the classical equations of motion and $\phi_\text{cl}(x,t)$ denotes the classical solution of the 1D wave equation corresponding to the initial conditions $\phi_0(x)$ and $\dot{\phi}_0(x)$. We have also used the fact that the operator $\cos(\phi)$ is diagonal in the $\phi$-representation. The solution $\phi_\text{cl}(t)$ can be explicitly constructed using d'Alembert's formula. After switching to dual-field representation using $K u \partial_x \theta = \dot{\phi}$, we get
\begin{eqnarray}
\langle \cos(2 \phi) \rangle &\sim& \int \mathcal{D}(\phi_0,\theta_0) \, \varrho_W(\phi_0,\theta_0) \, \cos \! \Big[ \phi_0(x-u t) \notag \\
&+& \phi_0(x+u t) + K \theta_0 (x+u t) - K \theta_0 (x-u t) \Big] \notag
\end{eqnarray}
Since in the initial state $\phi$ is pinned at $\phi_0=0$, we factor out the $\phi$ dependent part of the integral, obtaining
\begin{equation}
m_s(t) \sim \big\langle \cos K (\theta(u t)-\theta(-u t)) \big\rangle_0 \ ,
\label{LLms}
\end{equation}
where the brackets with the index $0$ denote the expectation value taken with respect to the initial state.
The r.h.s. of Eq. (\ref{LLms}) can be estimated within a semiclassical analysis, where the ground state of the Luttinger Hamiltonian (\ref{eq:llhamiltonian}) with an additional mass-term $\sim \Delta_s \phi^2$ is used as the initial state. This finally leads to
\begin{eqnarray}
m_s(t) &\sim&  \exp \, -\frac{K^2}{2} \left\langle (\theta(ut)-\theta(-ut) )^2 \right\rangle_0  \notag \\
&\sim& \exp \, -K \int_0^\Lambda dq \, \frac{\sqrt{q^2+\Delta_s^2/u^2}}{q^2} \sin^2(q u t)   \notag \\
&\xrightarrow{\Delta_s t\gg }& \exp(- \pi K \Delta_s t/2) \, ,
\label{LLresult}
\end{eqnarray}
where $\Delta_s$ again denotes the gap of the initial state\footnote{Iucci and Cazalilla \cite{iucci-2009} have recently derived an equivalent result.}. 
In contrast with the empirical rule (\ref{eq:af2}) for the XXZ model, the Luttinger model, being a continuum theory, does not reproduce oscillations. The non-oscillatory relaxation in (\ref{LLresult}) is characterized by a relaxation time inversely proportional to the gap and to the Luttinger parameter, $\tau=\frac{2}{\pi K\gap}$, a behavior identical to the conformal field theory result (\ref{eq:conformallocal}) and similarly observed in the numerical calculation for the quench in the XXZ model. However, the algebraic prefactor present in the case of the XXZ model (\ref{eq:af2}) and the spin-density-wave initial state under the XX Hamiltonian (\ref{eq:mssdw}) is not present in this treatment of the Luttinger model. Since the Luttinger model includes the XX limit at $K=1$, we conclude that the missing algebraic prefactor is a shortcoming of the initial state, which has been approximated as the ground state of the Klein-Gordon model (\ref{eq:kleingordon}). More accurate results could provide a treatment using the sine-Gordon Hamiltonian, which as we shall see in the next section strongly complicates the problem.

\section{Gapped theory -- the sine-Gordon model}
\label{sec:sg}

In this section we analyze the quench in the sine-Gordon model (see appendix \ref{sec:magnetismtheo}),
\beq
H_{SG} &=& \frac{1}{2\pi}\int
dx[uK(\pi\Pi(x))^{2}+\frac{u}{K}(\nabla\phi(x))^{2}]\notag\\
&-&\frac{2J\D}{(2\pi\alpha)^{2}}\int dx \cos(4\phi(x)),
\eeq
as a possible continuum approach to the quantum quench the XXZ
chain for $\Delta >1$. In what follows we use the boundary-state formalism as a convenient tool for describing the
non-equilibrium problem \cite{gritsev-2007}. In this formalism the initial state, which is not the
eigenstate of the Hamiltonian, can be thought of as a special
superposition of pairs of eigenmodes of the quantum Hamiltonian with
opposite momenta, which sums up into
a {\it squeezed} state of eigenmodes \cite{ghoshal-1994}. 
Unlike for the gapless Luttinger-liquid theory, we cannot present a full solution of the dynamics. Possible directions to be followed in future are pointed out.

Since the sine-Gordon model has relativistic (Lorentz) invariance,  we can
exchange the time and space directions $x\leftrightarrow t$ and
consider the following boundary-in-time Hamiltonian (using more
conventional notation $x$ again for the imaginary time direction)
\beq
H &=& \frac{1}{2\pi}\int
dx[uK(\pi\Pi(x))^{2}+\frac{u}{K}(\nabla\phi(x))^{2}]\nonumber\\
&-&\frac{2J}{(2\pi\alpha)^{2}}\int dx\{ 
\D\cos(4\phi(x)))\theta(x)\nonumber\\&+&\Delta_0\cos(2\phi(x))\delta(x)\}\,,
\eeq
where $\theta(x)$ is a theta-function and $\delta(x)$ takes care of
the initial condition. In order to implement the N\'eel state as an initial condition we send $\Delta_0\rightarrow\infty$ which
corresponds to the Dirichlet boundary (initial) condition. This
boundary (initial) {\it condition} formulation can be reformulated
in a boundary-{\it state} formalism of the boundary sine-Gordon model
(bSG). The initial condition is
expressed as a {\it squeezed state} of bulk degrees of freedom. We
note that for noninteracting particles or Luttinger liquid this
correspondence can be seen directly. Since for $K<1/2$ there are
only solitons and antisolitons in the spectrum (repulsive regime of
the sine-Gordon model), we obtain the boundary state in the following
form
\beq
|B(t=0)\rangle_{D}={\cal N}\exp\left[\int
K^{ab}_{D}(\theta)A^{\dag}_{a}(\theta)A^{\dag}_{b}(-\theta)\right]|0\rangle\,.
\eeq
Here $A^{\dag}_{a,b}(\theta)$ is an operator of creation of the
soliton ($a$) or antisoliton ($b$) and $K^{ab}_{D}(\theta)$ is a reflection
matrix of soliton-antisoliton pair corresponding to the Dirichlet
boundary condition. The rapidity $\theta$ is related to
the momentum $P=M_{s}\sinh\theta$ and energy
$E=M_{s}\cosh\theta$, where the soliton mass $M_s$ is given by \cite{zamalodchikov-1995}
\beq
M_{s}=\left(\frac{\frac{J\D}{2\pi\a^2}\Gamma(1-\frac{\beta^{2}}{8\pi})
}{\Gamma(\frac{\beta^{2}}{8\pi})}\right)^{\frac{1}{2-2\frac{\beta^{2}}{8\pi}}}
\frac{2\Gamma(\frac{\xi}{2})}{\sqrt{\pi}\Gamma(\frac{1}{2}+\frac{\xi}{2})}\,,
\eeq
where we define $\xi =\beta^{2}/(1-\beta^{2})$.

The evolution is trivial in the soliton basis, because the bulk Hamiltonian
is diagonal in soliton-antisoliton operators,
\beq
|B(t>0)\rangle_{D}&=&{\cal N}\exp\left(\int
K^{ab}_{D}(\theta,t)A^{\dag}_{a}(\theta)A^{\dag}_{b}(-\theta)\right)|0\rangle\nonumber\\
K^{ab}_{D}(\theta,t)&=&K^{ab}_{D}(\theta)\exp(2it
M_{s}\cosh(\theta))\,.
\eeq

In the boundary state formulation the evolution of the magnetization
is equivalent to the computation of the following quantity
\beq
m_{s}(t)=\langle B(t)|\cos(2\phi(0))|B(t)\rangle \label{eq:boundaryms}\,.
\eeq

In general the squeezed state represented by the boundary state
$|B(t)\rangle$ should be expanded as a series in powers of the reflection matrices. This produces
multiple dynamical processes which include solitons and
antisolitons. Multi-particle expectation values of the operators,
like $\cos(2\phi)$, are called form-factors. To compute the
correlation functions in the massive theories at equilibrium,
only a small number of lowest form-factor contributions is
necessary. However, our evaluation of the lowest order contributions in our case provided results contradictory to the numerical simulations. The reason for this will be found in the spectral analysis of the next section, which hints that not only
soliton-antisoliton form-factors are important (which is the
case for the spectral function of the sine-Gordon model for small
energies), but also multiple processes, which include energies well
above the spectral gap (soliton mass), are necessary to be
considered. Technically, the problem of inclusion of multi-soliton
form-factors is rather difficult. The $\theta$-integrals
corresponding to evaluation of different multi-particle
contributions become even more complicated because of the reflection
matrices $K^{ab}_{D}(\theta)$.

A possible alternative approach to this form-factor evaluation could be a
resummation of the leading divergencies of the scattering processes
in the presence of a boundary state. Since the ultra-violet energies give an
important contribution in our problem, one can try to proceed by
considering the logarithm of the one- or (two-) point function and to
sum the leading contributions as proposed previously \cite{smirnov-1990,lesage-1997,babujian-2004,takacs-2008}.
However, the complexity of the boundary reflection matrix does not
allow to realize this program. We hope to return to this problem in
future.

In view of the high complexity of the boundary-state formalism it may be worthwhile to establish phenomenological analogies between non-equilibrium dynamics and equilibrium dynamics of the sine-Gordon model. This would be useful since for calculating dynamical structure factors a powerful machinery has been developed over the last decades. 
Arguing that the initial state can be described by a thermal ensemble (with some effective temperature considered as a
fitting parameter) instead of the boundary state, we can relate the dynamics of the magnetic order parameter to the
two-point function, 
\beq
\ms(t)\sim\langle\cos(2\phi(t))\cos(2\phi(0))\rangle\label{eq:mstwopoint}\,,
\eeq
where the average is taken over some thermal ensemble characterized by temperature $T_{eff}$. The operation of $\cos(2\phi(0))$ onto the thermal state is a possibility to introduce some magnetic order, or, stated otherwhise, to establish an analogy to Eq. (\ref{eq:boundaryms}); $\cos(2\phi(t))$ acting on the thermal state is a way to mimic a boundary-in-time state. We note that such approach has been successfully applied for studying dynamics of a non-local observable in quench in the quantum Ising chain \cite{rossini-2009}.
The dynamics of the two-point function (\ref{eq:mstwopoint}) is separated into two regimes: large-temperatures $T_{eff}\gg M_{s}$ and low-temperatures $T_{eff}\ll M$. It is
known that for large energies (UV) massive models, like the
sine-Gordon model, have the conformal filed theory asymptotics.
Therefore, in the large-temperature regime the behavior of the
correlation functions should be the same as in the high-temperature
limit of the corresponding  conformal field theory. Hence, for
$T_{eff}\gg M_{s}$ the large-time asymptotics of the correlation
function is given by an exponential decay
\beq
m_{s}(t)\sim \exp[-\pi T_{eff} \frac{K}{2} t].
\eeq
This conformal field theory
behavior is universal also for the gapless phase, where, at least in some regimes of weakly magnetized initial states, setting $T_{eff}=\gap$ this behavior is a good first approximation of the dynamics of the order parameter in the quench problem (see Table \ref{tab:results}).
However, in the gapped phase we cannot find a reasonable way to define $T_{eff}$. For example, the temperatures corresponding to the Boltzmann ensembles used in the following section do not reproduce at all the numerical findings.

In the other regime, $T_{eff}\ll M_s$, the structure of the massive
theory is important. In this case the leading order behavior comes from the
zero-momentum exchange processes and depends on the structure of
scattering matrix $S(0)$ in this limit. Resummation of the
kinematical singularities leads again to the exponential decay for
the two-point correlation function \cite{altshuler-2006}, in agreement with a
quasi-classical formula from Ref. \cite{sachdev-1997}. Implementing results of
\cite{altshuler-2006} to our situation we obtain
\beq
m_{s}\sim \exp[-T_{eff}e^{-M_{s}/T_{eff}}t],
\eeq
where the proportionality coefficient depends on the power of
$M_{s}$. Such behavior however is in disagreement with our numerical findings, where in the limit of large $M_s$ (large $\Delta$) we find a decay rate proportional to $\Delta^{-2}$.

We conclude that although the sine-Gordon is a valuable candidate for describing the dynamics following a quantum quench in the XXZ model, the evaluation of the corresponding form-factors is difficult and demands further efforts. A relation of the coherent dynamics of the order parameter to dynamical structure factors, circumventing this problem, is not straightforward to be established. 

\renewcommand{\d}{\ensuremath{\delta}}

\section{Spectral analysis}
\label{sec:spectral}
\begin{figure*}
\centering
\includegraphics[height=\figw3\textwidth,angle=0]{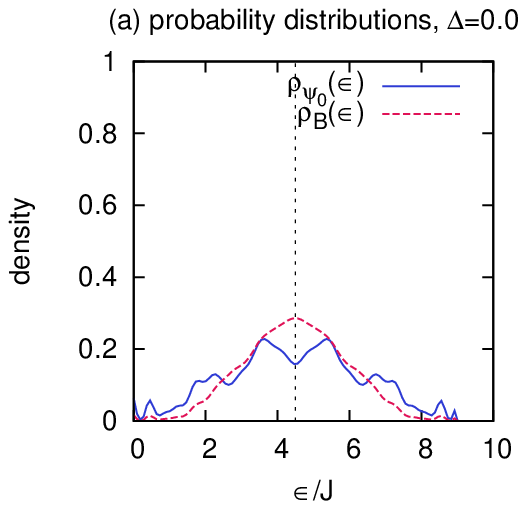}
\includegraphics[height=\figw3\textwidth,angle=0]{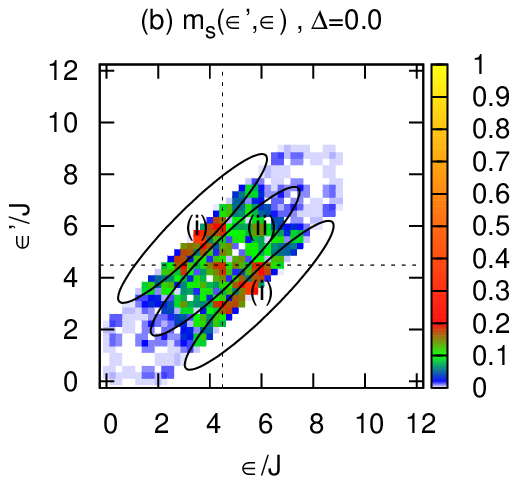}
\includegraphics[height=\figw3\textwidth,angle=0]{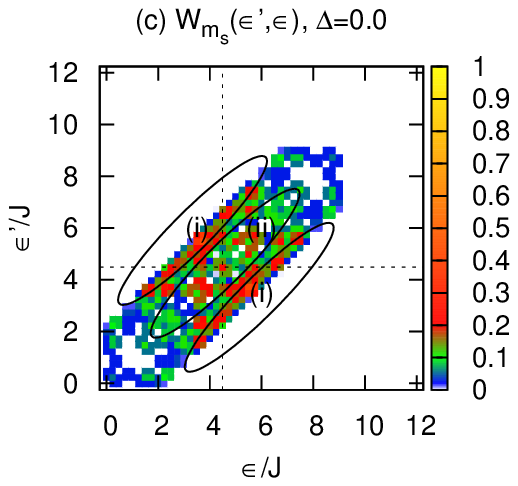}
\includegraphics[height=\figw3\textwidth,angle=0]{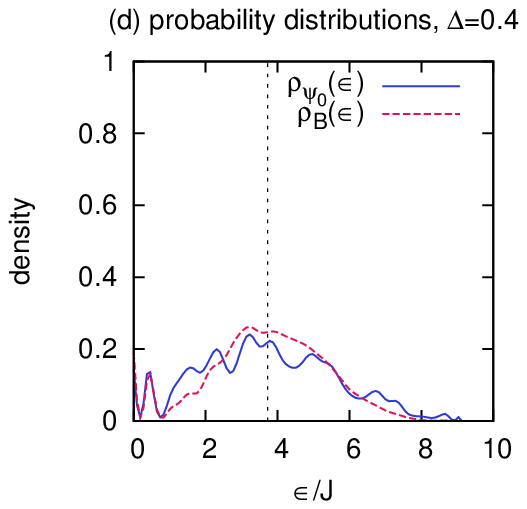}
\includegraphics[height=\figw3\textwidth,angle=0]{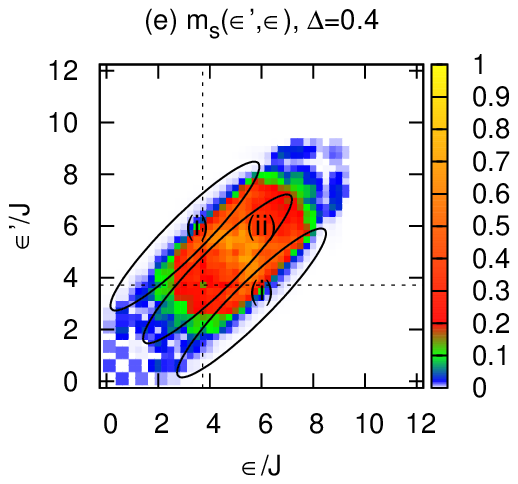}
\includegraphics[height=\figw3\textwidth,angle=0]{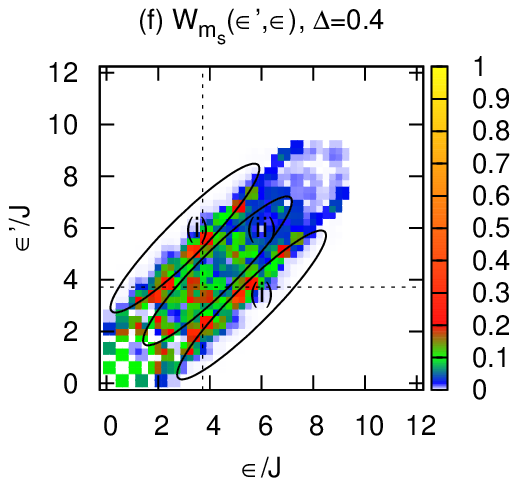}
\includegraphics[height=\figw3\textwidth,angle=0]{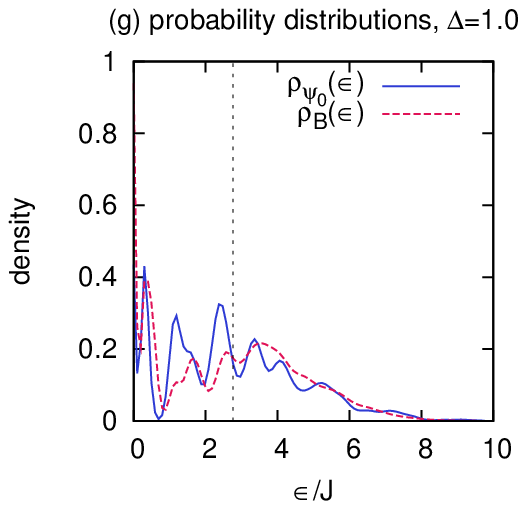}
\includegraphics[height=\figw3\textwidth,angle=0]{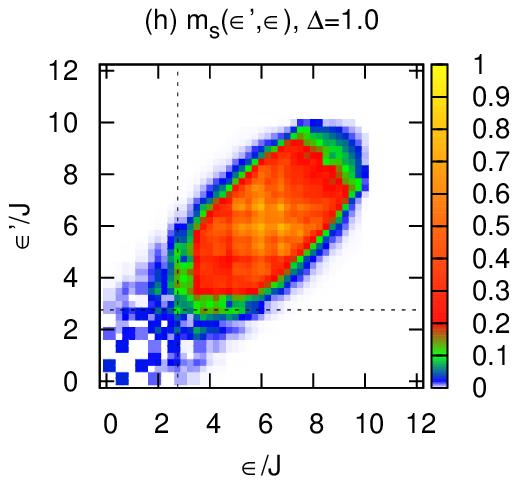}
\includegraphics[height=\figw3\textwidth,angle=0]{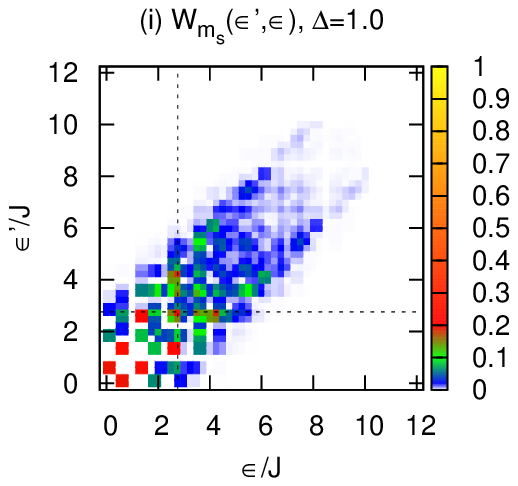}
\includegraphics[height=\figw3\textwidth,angle=0]{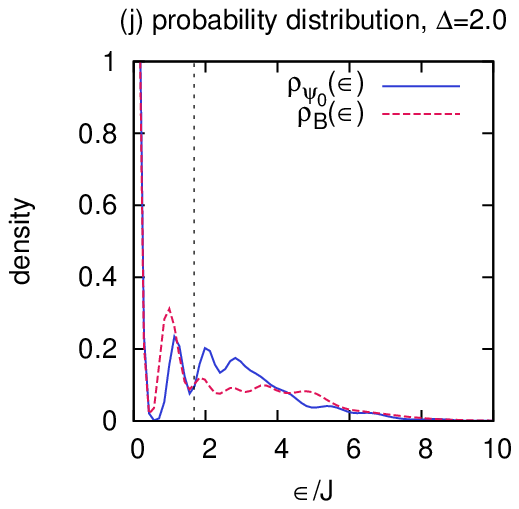}
\includegraphics[height=\figw3\textwidth,angle=0]{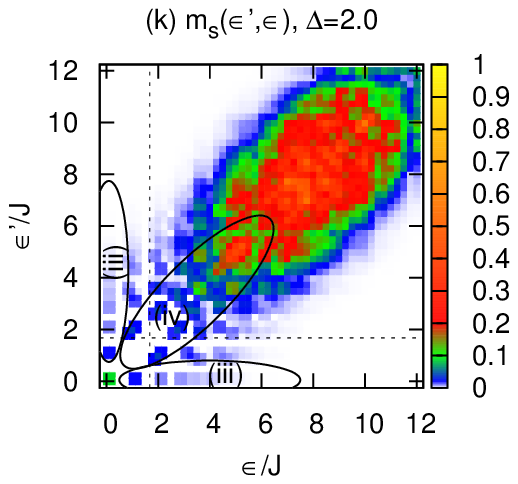}
\includegraphics[height=\figw3\textwidth,angle=0]{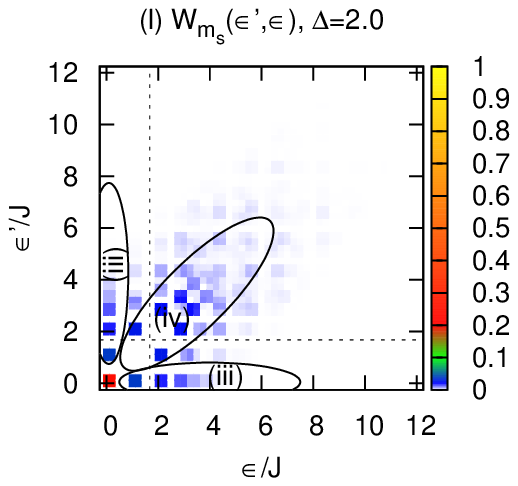}
\includegraphics[height=\figw3\textwidth,angle=0]{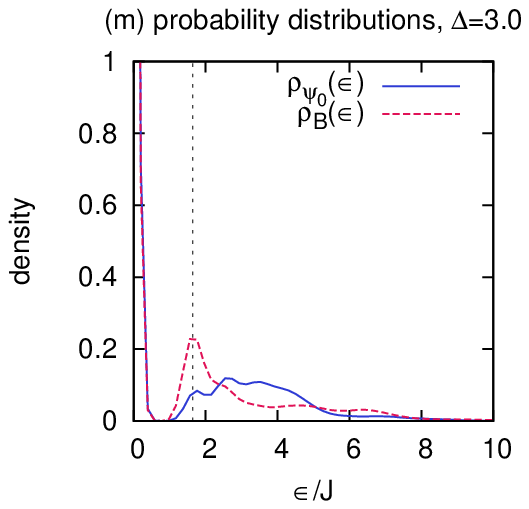}
\includegraphics[height=\figw3\textwidth,angle=0]{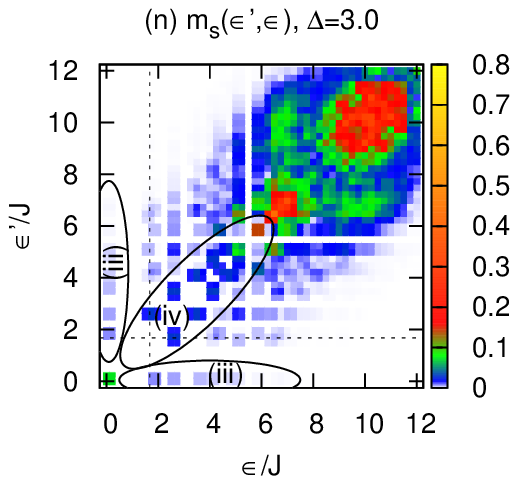}
\includegraphics[height=\figw3\textwidth,angle=0]{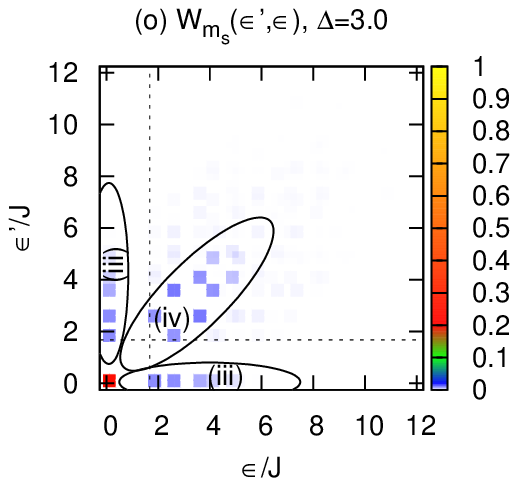}
\caption{\label{fig:spectral}Analysis of the spectrum of the XXZ chain for a system of 14 sites with periodic boundary conditions. Dotted lines mark the position of the energy of the system, $E=\bra{\psi_0}{H}\ket{\psi_0}$. See text for the description of the regions marked $(i)$-$(iv)$.}
\end{figure*}
\begin{figure*}
\centering
\includegraphics[width=0.98\textwidth,angle=0]{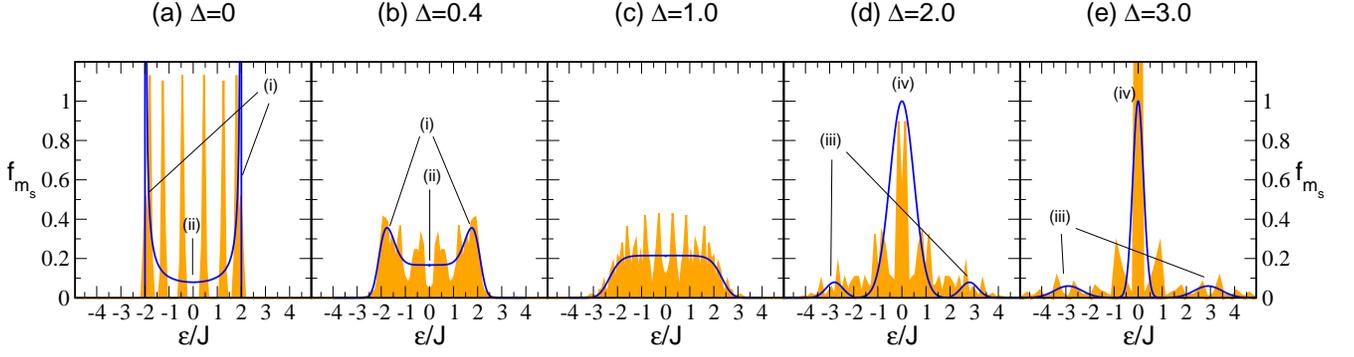}
\caption{\label{fig:integrated}The frequency distribution of the staggered magnetization,  $f_\ms(\e)$, for a system of 14 sites with periodic boundary conditions. Histograms resolve individual peaks, solid lines correspond to (non-unique) smoothened distributions emphasizing separable contributions. For $\D=0$ the exact solution in the thermodynamic limit (\ref{eq:xxdist}) is drawn instead.}
\end{figure*}

For a deeper understanding of the relaxation dynamics, it is useful to consider the problem in energy space. The idea is to associate properties of the spectrum of the Hamiltonian to the dynamical phenomena observed in the simulation of the time evolution and to clarify the possibility of separating energy scales -- a question which is especially important for improving  analytical descriptions of the non-equilibrium dynamics.

Using the Lehmann representation, the time evolution of an operator $O$ takes the form of a Fourier transform over the eigenlevels of the Hamiltonian,
\beq
\av{O(t)}=\sum_{n,m}e^{-it(E_m-E_n)}\braket{\psi_0}{m}\bra{m}O\ket{n}\braket{n}{\psi_0}\notag\,.
\eeq
For a more convenient continuum description, we introduce the \textit{quenched probability distribution},
\beq
\rho_{\psi_0}(\e)=\sum_n\d(\e + E_0- \e_n )|\braket{n}{\psi_0}|^2\,,
\eeq
which determines the properties of the stationary state at $t\rightarrow\infty$ \cite{manmana-2007,silva-2008,polkovnikov-2008,reimann-2008} (the frequencies $\e$ are shifted by $E_0$ -- the ground-state energy of $H$). It can be compared to the \textit{thermal} (Boltzmann) distribution of the grand canonical ensemble,
\beq
\rho_B(\e)=\frac{1}{\mathcal{N}}\sum_n\d(\e + E_0 - \e_n )e^{-\e_n/T}\,,
\eeq
where the temperature is set by the energy of the initial state, $\int d\e\e\rho_B(\e)=\bra{\psi_0}H\ket{\psi_0}$. In general, it is known that the thermal distribution can deviate strongly from the quenched distribution \cite{rigol-2007,manmana-2007,silva-2008,polkovnikov-2008,reimann-2008}, which leads to the phenomena of {\it absence of thermalization}. Here the thermal distributions are used only as a reference and questions in connection with thermalization phenomena will not be investigated.
While the quenched probability distribution, $\rho_{\psi_0}$, captures the effect of the initial state, the distribution of the expectation value,
\beq
O(\e',\e)=\sum_{n,m}\d(\e'  + E_0- \e_m)\d(\e + E_0 - \e_n)\bra{m}O\ket{n}\notag\,,
\eeq
reflects the specific spectral properties of the given observable. 
$\rho_{\psi_0}$ and $O(\e',\e)$ provide the contributions to the weighted expectation value,
\beq
W_{0}(\e',\e)=\sum_{n,m}\d(\e'  + &&E_0- \e_m)\d(\e + E_0 - \e_n)\notag\\&&\times\braket{\psi_0}{m}\bra{m}O\ket{n}\braket{n}{\psi_0}\,,
\eeq
which, via the distribution function
\beq
f_O(\e)=\int d \e' W_O(\e+\e',\e')\,,
\eeq
represents the dynamics of the observable in frequency space,
\beq
O(t)=\int d\e e^{-i\e t}f_O(\e)\,.
\eeq

The spectral properties of the XXZ chain prepared in the N\'eel state, $\ket{\psi_0}=|\ua\da\ua \cdots \da\ua\rangle$, with the staggered magnetization as the observable, $O=\ms$, are calculated by means of exact diagonalizations for small system sizes. Fig. \ref{fig:spectral} displays the results for a chain of length $N=14$. The small system size results in strongly peaked distributions, but we have made sure that the qualitative features we extract in the following analysis are stable against variations of the system size, both towards larger, $N=18$, and smaller, $N=10$, values.  Quantitative information cannot be extracted from this simple analysis, but this might be possible when going to larger system sizes by means of more involved techniques, such as the Lanczos method \cite{roux-2009}.

In the non-interacting limit ($\D=0$), where the Hamiltonian has a free-fermion representation as discussed in section \ref{sec:xx}, all distributions are centro-symmetric about $\e=E_0$. The quenched distribution $\rho_{\psi_0}(\e)$ exhibits peaks at $\e-E_0=\pm J$, which are not present in the thermal distribution. From the discussion in section \ref{sec:xx} and the finite size-study [Fig. \ref{fig:spectral}(a-c)] we can separate two contributions to $W_\ms(\e',\e)$:
\renewcommand{\labelenumi}{\alph{enumi}}
\begin{enumerate}
 \item[(i)] \textit{Free-fermion band edges.}  The distribution has maxima and sharp cutoffs at the edges $\e-\e'=\pm 2J$. 
\item[(ii)] \textit{Low frequency.} The contributions from the areas at $\e-\e'\ll J$ are weaker than the band-edge contributions $(i)$.
\end{enumerate}
For small but finite anisotropies, as exemplified by the results for $\Delta=0.4$ [Fig. \ref{fig:spectral}(d-f)], the main features  of the $\Delta=0$ distributions are still present. However, the distributions lose their reflection symmetry and the weight is shifted towards lower energies. In addition to this asymmetry, the cutoff of the distribution of $\ms$ is no longer absolutely sharp at the band edge. This can be understood as an effect of the breakdown of the free-fermion quasi particle description of the Hamiltonian at $\D\neq0$.

For large anisotropies in the gapped regime [e.g. $\Delta=3$, Fig. \ref{fig:spectral}(m-o)], the quenched probability distributions have a single peak at $\e=0$ and a continuum above the gap, which is rather flat in comparison with the thermal distribution. The expectation values of the observable $\ms(\e',\e)$ are spread over a  large energy region, however, the relevant weighted expectation values $W_\ms(\e',\e)$ receive contributions  of only three different types, all of which are located at low energies:
\renewcommand{\labelenumi}{\roman{enumi}.}
\begin{enumerate}
 \item[(iii)]\textit{Continuum above the gap}. The contributions form $\bra{m}\ms\ket{n}$, $n\neq0,m\neq0$ form a rather continuous distribution with frequencies far below the gap $\e-\e'\ll J\D$. 
 \item[(iv)] \textit{Off-diagonal contributions involving the ground state.} Elements $\bra{0}\ms\ket{n}$, $n\neq0$, result in frequencies equal to or larger than the gap.
\item[(v)] \textit{Ground state.} An isolated peak is located at the ground state energy of H ($\e=\e'=0$).
\end{enumerate}

For lower $\D$ this picture remains basically valid. However, in Fig. \ref{fig:spectral}(j-l), where the results for $\D=2$ are shown, the width of the contribution $(iv)$ becomes of the order of the gap. For even lower anisotropies, e.g. $\D=1$, separation of the contributions is no longer possible [Fig. \ref{fig:spectral}(g-i)].

The effects of these contributions on the frequency distribution $f_{\ms}(\e)$ is shown in Fig. \ref{fig:integrated}. The characteristics of time evolution can now be inferred from the properties of the Fourier transform of the distribution $f_{\ms}(\e)$. The algebraically decaying oscillations at $\D=0$ (\ref{eq:msxx}) are a consequence of the step-like shape of $f_{\ms}(\e)$, which is a consequence of the properties of the contribution $(i)$. For $\D>0$ the smearing  of the edge is the cause of the exponential decay of the oscillations. The finite width of the edge in Fig. \ref{fig:integrated}(b) is consistent with the relaxation time of the oscillations (Fig. \ref{fig:relaxtime_neel}). At $\D>1$ there are also contributions at large frequencies originating  from contributions of type $(iv)$, resulting in rather broad peaks in $f_{\ms}(\e)$, which lead to the quickly decaying oscillations seen in Fig. \ref{fig:ms_strong}. The low-frequency contributions $(iii)$ are reflected in a peak of $f_{\ms}(\e)$ around $\e=0$, whose width corresponds to the relaxation time $\tone$ of the non-oscillatory decay (Fig. \ref{fig:relaxtime_neel}). 

The isolated peak $(v)$ at zero energy is irrelevant for the dynamics. It would correspond to a finite asymptotic value at $t\rightarrow\infty$, which apparently vanishes in the thermodynamic limit.

In summary, on the basis of the analysis of the frequency distributions, we can now draw a qualitative crossover picture from the oscillatory to the non-oscillatory behavior as $\D$ varies. Approaching the isotropic point from small values of $\D$, the band edges $(i)$ are smeared out, leading to a decreasing relaxation time $\ttwo$. Starting from large values of $\D$, the low-frequency peak merges into the homogeneous distribution upon approaching the isotropic point. Hence the characteristics of $(i)$ and $(iii)$ contributions, dominating at small (respectively large) values of $\D$, are both lost at intermediate values of $\D$, where the interplay of all energy scales apparently leads to a non-generic dynamical relaxation of the order parameter. We also note that, even in the regime $\D\gg1$, where the initial state is rather close to the ground state of the Hamiltonian, the relevant part of the spectrum located above the gap is a multi-particle continuum, difficult to be treated analytically.

\section{Conclusions and Outlook}
\label{sec:conclusion}

We have analyzed the dynamics of the staggered magnetization in quantum spin chains following a quantum quench, considering various antiferomagnetically ordered initial states by using a number of complementary numerical and analytical approaches.
In the numerical MPS study we have essentially found three types of relaxation dynamics for the order parameter:
(i) For highly ordered initial states ($\D_0\gg1$) and sufficiently small anisotropy parameters of the Hamiltonian at $t>0$ there are Rabi oscillations, which dephase exponentially in time away from the XX limit; (ii) for strong anisotropies ($\D\gg1$) there is an exponential decay and the relaxation time scales as $\tau\propto\D^{2}$; (iii) for initial states close to the phase transition we found evidence for algebraic corrections to the exponential decay.  There is a crossover phenomenon between oscillatory (small $\D$) and non-oscillatory dynamics (large $\D$), but no clear point of transition can be identified. Either both types of dynamics superimpose on each other, as it is the case for $\D_0\gtrsim1$, or both vanish in an extended transition regime, in which a non-generic behaviour is found (case of $\D_0\gg1$).

We have shown that a precise description of the Rabi oscillations (i) is  possible only when the full spectrum is taken into account. Therefore mean-field as well as low-energy approaches lead to incomplete results -- an analytical treatment of this type of dynamics is feasible only by novel approaches. It has become clear that quasi-particles relevant for Rabi-oscillations are located at the band edges. In contrast to equilibrium properties, where many-body effects can be incorporated into the relevant linearizable Fermi-level excitations within the bosonization formalism, it is not obvious how to treat interactions in combination with the quadratic dispersion relation at the edges of the band. A treatment along the lines of previously developed concepts dealing with non-linear effects in dynamical phenomena \cite{bettelheim-2006,imambekov-2009} might be a possible solution to this problem. 

While the exponential decay at large anisotropies (ii) appears to be a rather generic behaviour and is also reproduced in the exactly solvable XZ model, the algebraic prefactor in front of the exponential law (iii) is a more intricate phenomenon. In the XX limit, where we approximated the initial state by a spin-density wave, such order-parameter dynamics are reproduced. 
However, standard field-theoretical approaches, such as conformal field theory or the description by Luttinger model adopted here, capture roughly the scaling of the corresponding relaxation time, but do miss the prefactor. Our results suggest that this is an consequence of the phenomenological description of the initial state in terms of a simple massive theory (Klein-Gordon). A more elaborate treatment of the initial state using the sine-Gordon model could resolve this deficiency of the field-theoretical descriptions.

The sine-Gordon as well as the original XXZ model is integrable. Analyzing the quench by using the integrability of these models amounts to evaluation of form factors of particles with non-trivial statistics. While for two or three particles this problem can be solved \cite{smirnov-1992,controzzi-2001}, we have found that non-equilibrium dynamics require the evaluation of higher order form factors -- a yet unsolved and highly complex problem. A promising approach is to use the structure of the Bethe-ansatz solution in combination with a numerical algorithm \cite{faribault-2009}.
 
Similarly to studies of other models \cite{cramer-2008,eckstein-2009} we find as a generic feature that relaxation times become small in the vicinity of the quantum critical point. However, the behaviour is far too rich to be attributed to a generic dynamical phase transition \cite{eckstein-2009}. In our analysis we have shown that effective descriptions by low-energy theories belonging to the universality class of the model can not capture all relevant processes. 
A sort of a dynamical phase transition occurs, however, in the mean-field description (a finite magnetization is found in the long-time limit for $\D>1$), which treats interaction terms as on an infinite-dimensional lattice.  This hints that the existence of a sort of dynamical critical behavior may, very much like for equilibrium phase transitions, depend on dimensionality. A first step towards understanding of the role of dimensionality is the study of non-equilibrium dynamics in infinite dimension, for example by using dynamical mean-field theory \cite{freericks-2006,eckstein-2009}. How to treat coherent dynamics in a two- or three-dimensional system is still an open question.

Finally, we have to mention that the Heisenberg chain is a simplified model, well suited for numerical and analytical investigations, but not necessarily appropriate for full description of experimental systems. Although in experiments with two-level atoms in optical lattices  behaviour similar to our theoretical prediction is observed \cite{trotzky-2008,trotzky-private}, the model has to be adjusted to provide an accurate description of the experimental results. For example, the effect of density fluctuations beyond the purely magnetic model needs to be investigated. Although the matrix product algorithm is an efficient method to study the relaxation dynamics, the numerically reachable times are fundamentally restricted by growing entanglement. The runaway time may become very small in  models more involved than spin-$\half$ chains, in particular when particle fluctuations need to be taken into account \cite{barthel-2008}. Recently, schemes have been proposed which allow to go beyond what is possible within the conventional matrix product algorithms \cite{white-2008,barthel-2009,hastings-2009,banuls-2009}. The main idea in these approaches is to calculate directly the dynamics of an observable rather than explicitly follow the evolution of the state. Another important aspect neglected here, but relevant in experiments, is temperature. How to efficiently include effects of finite temperature within a time-dependent matrix product algorithm is a yet unsolved problem \cite{barthel-2009,verstraete-2004}.

\section{Acknowledgements}
We thank D. Baeriswyl, I. Bloch, C. Kollath, M. Lukin, S. R. Manmana, M. Menteshashvili, A. M. Rey, M. Ringel, A. Silva, S. Trotzky and W. Zwerger for fruitful discussions. Part of our simulations are based on the ALPS library\cite{alps-2007}. This work was supported by SNF (P.B. and V.G), DFG - FOR 801 and the Feinberg Graduate School (M.P.), US Israel Binational Science Foundation (E.A. and E.D.) and AFOSR, CUA, DARPA, MURI, NSF DMR-0705472 (V.G. and E.D.).

\appendix

\section{Quantum magnetism in optical lattices}
\label{sec:magnetismexp}
The underlying model for realizing quantum magnetism in optical lattices is a single-band Hubbard model,
\beq
H&=&\sum_{ij\s}\left\{t_{ij\s }a_{i\s}^\dagger a_{j\s}+ \mbox{H.c.}\right\} +
\sum_i U_{\ua\da}n_{i\ua}n_{i\da}\notag\\&+&\sum_{i\s}\frac{U_{\s\s}}{2}(n_{i\s}-1)n_{i\s}-\sum_i\mu_{i\s}n_{i\s}\,.
\label{eq:bosehubbard}
\eeq
$a_{i\s}^\dagger$ are the creation operators of a particle in a Wannier state of type $\s$ at site $i$, satisfying bosonic commutation relations. $n_{i\s}=a_{i\s}^\dagger a_{i\s}$ are the corresponding occupation numbers. The internal degree of freedom $\s$ represents typically a hyperfine state and can be identified with (pseudo) spin-$\half$, $\s=\ua,\da$. The hopping integral $t_{ij\s}$ and the interaction parameters $U_{\s\s'}$ depend on the geometry and the depth of the lattice and can be expressed in terms of overlaps of the Wannier orbitals. If, for concreteness, we consider a periodic, spin-dependent lattice potential with an isotropic spacing $a$, 
\beq
V_\s(\br)=\sum_{\mu=x,y,z}V_{\mu\s}\sin^2 Kr_\mu\,,
\eeq
with $K=\frac{2\pi}{a}$, the recoil energy needs to be much smaller than the lattice depth, $E_r=\frac{\hbar^2K^2}{2m}\ll V_{\mu\s}$, so that the atoms are in the lowest harmonic level and single-band description (\ref{eq:bosehubbard}) is valid. In reality, the gaussian shape of the laser beams introduces  inhomogeneity in the lattice depth in addition to the harmonic trapping potential $\mu_i$. The superposition of polarized laser beams generates spin-dependent potentials \cite{mandel-2003a,jaksch-1999}. In general, the orbitals extend only over short distances and the hopping can be restricted to nearest neighbors, $t_{ij\s}=t_{\mu\s}$, if $\br_i-\br_j={\bf e}_{\mu}a$, where \cite{duan-2003}
\beq
t_{\mu\s
}&\approx& \left( 4/\sqrt{\pi }\right)E_{r}^{1/4}\left( V_{\mu\s
}\right) ^{3/4}\exp [-2(V_{\mu \s }/E_{r})^{1/2}]\,,
\eeq
\beq
U_{\s
\s' }&\approx& (8/\pi )^{1/2}(ka_{s\s \s' })(E_{r}%
\overline{V}_{x\s \s' }\overline{V}_{y\s \s' }%
\overline{V}_{z\s \s' })^{1/4}\,.
\eeq
Here, $\overline{V}_{\mu
\s -\s }=4V_{\mu \ua }V_{\mu \da }/(V_{\mu \ua
}^{1/2}+V_{\mu \da }^{1/2})^{2}$, is the spin-average potential in
each direction, $\overline{V}_{\mu
\s \s }=V_{\mu\s}$, and $a_{s\s\s' }$ is the s-wave scattering length between atoms of
spin $\s$ and $\s'$.

In the case of strong on-site repulsion, $t_{\mu\s }\ll U_{\s\s' }$, and integer filling, the system is in the Mott phase, where occupation number fluctuations are essentially suppressed. In this case, the effective basis contains locally only singly occupied spin-up $\ket{\ua}$ or -down $\ket{\da}$ states (for  the sake of simplicity we choose $\left\langle n_{i\ua
}\right\rangle +\left\langle n_{i\da }\right\rangle = 1$, although higher occupation numbers are also possible). In this subspace, neglecting terms of order $t_{ij}^4/U_{\s\s'}^3$, the Hubbard model (\ref{eq:bosehubbard}) can be mapped onto a spin-$\half$ Heisenberg model (XXZ model) \cite{anderson-1959,schrieffer-1966,chao-1977,fazekas-book},
\beq
 H_\text{XXZ} =  \sum_{ij} \left\{J_{\bot}^{ij} (S_i^x S_{j}^x + S_i^y S_{j}^y) + J_{z}^{ij}  S_i^z S_{j}^z \right\}\,.
\label{eq:xxzhamiltonian}
\eeq
The superexchange interaction constants $J_{z}^{ij}$ and $J_{\bot}^{ij}$  are given by
\beq
J_{z}^{ij}&=&\frac{2t_{ij\ua }^{2}+2t_{ij\da }^{2}}{U_{\ua \da }}-\frac{4t_{ij\ua }^{2}}{U_{\ua\ua }}-\frac{4t_{ij\da }^{2}}{U_{\da\da }}\,,\\
J_{\bot}^{ij}&=&-\frac{4t_{ij\ua }t_{ij\da }}{U_{\ua \da }}.
\label{eq:superexchangecouplings}
\eeq

An analogous treatment can be carried out for the fermionic Hubbard model. In the resulting magnetic Hamiltonian (\ref{eq:xxzhamiltonian}), $J_{\perp }$ has the opposite sign compared to Eq. (\ref{eq:superexchangecouplings}) and in the expression for $J_{z}$ the last two terms are absent since double occupancy is forbidden by the Fermi statistics.

For appropriately chosen lattice and interaction parameters, the anisotropy of the spin exchange, $\Delta^{ij}=J_{z}^{ij}/J_{\bot}^{ij}$, is tunable to a large extent. For example, in the bosonic case with symmetric on-site repulsions, a ferromagnet with possible easy-axis anisotropy, $\D^{ij}=\frac{1}{2}(\frac{t_{ij\ua}}{t_{ij\da}}+\frac{t_{ij\da}}{t_{ij\ua}})\geq1$, is realized. In addition, as demonstrated recently \cite{trotzky-2008}, double-well potentials can be used to change the sign of the exchange interactions from ferro- ($J_{z}^{ij}<0$) to antiferromagnetic ($J_{z}^{ij}>0$) \cite{barthel-2008}.

Although the Heisenberg Hamiltonian (\ref{eq:xxzhamiltonian}) is a good first approximation to strongly interacting two-component Bose-gases in optical lattices, we note that the measurements of Trotzky \etal~\cite{trotzky-2008} clearly show the limitations of the purely magnetic picture.  The strong repulsion leads to superexchange interaction which reaches the order of currently realistic temperatures, $J/k_B\sim10^{-9}K$, and in non-equilibrium experiments the dynamics slow down, so that effects of inhomogeneous laser beams become strong. For larger tunnelings density fluctuations are important and introduce an additional higher frequency; excitations to higher Bloch-bands may also become possible. We conclude that, although the experimental progress looks promising, further improvements in experimental setups are still needed in order to produce a clean realization of a quantum magnet.

\section{Equilibrium properties of the XXZ model in one dimension}
\label{sec:magnetismtheo}
In a spatially anisotropic optical lattice the Heisenberg chain, a paradigm in the theory of magnetism and strongly correlated systems in general, can be realized experimentally as proposed in the preceding Section. Here we give an overview of the equilibrium phases of the Heisenberg chain, focusing on antiferromagnetic exchange interactions. At the same time, the important concepts and notations to be used in the ensuing discussion of the non-equilibrium problem are introduced.

The one-dimensional spin-$\frac{1}{2}$ Heisenberg chain, 
\begin{equation}
H = J \sum_j \left\{ S_j^x S_{j+1}^x + S_j^y S_{j+1}^y + \Delta  S_j^z S_{j+1}^z \right\}\,,
\label{eq:xxzhamiltonianapp}
\end{equation}
is integrable -- the eigenstates and an infinite number of conserved operators can be obtained using the Bethe ansatz \cite{bethe-1931,cloizeaux-1966,yang-1966a,yang-1966b,yang-1966c}. A number of equilibrium properties can be exactly computed for the Bethe wave function -- examples are the energy and momentum of low-lying states, or local observables such as the staggered magnetization \cite{baxter-1973,baxter-book}. For some specific cases, non-local properties can also be calculated analytically \cite{korepin-book,kitanine-2009} or by means of a combination of the Bethe ansatz with numerical algorithms \cite{faribault-2009,caux-2005,caux-2005b}. A simplified insight into the physics of the Heisenberg chain can be gained from a continuum description via the bosonization  technique \cite{affleck-1988,giamarchi-book}. Here, results from both approaches, Bethe ansatz and bosonization, will be presented.

\begin{figure}[t]
\centering
 \centering
 \includegraphics[width=0.86\columnwidth,angle=0]{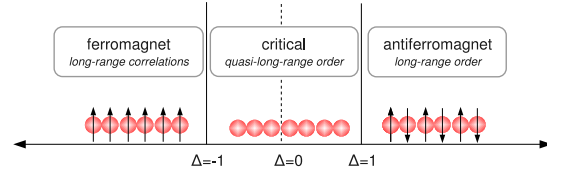}
 \caption{\label{fig:xxzphasediagram}The ground-state phase diagram of the XXZ model in one dimension.}
\end{figure}

The ground-state phase diagram of the XXZ model is represented in Fig. \ref{fig:xxzphasediagram}. Without loss of generality the coupling $J$ can be considered to be positive and the phases are simply characterized by $\D$. The long-range ordered antiferromagnetic phase for Ising-like anisotropies $\Delta>1$ exhibits a spectral gap. In the easy-plane regime $|\Delta|\leq1$, a critical gapless phase is found. The phase for $\Delta<-1$ is ferromagnetically ordered.

A useful equivalent representation of (\ref{eq:xxzhamiltonianoned}) is a  model of interacting spinless fermions,
\begin{eqnarray}
H_{XXZ} = \frac{J}{2} \sum_j\left\{ c^\dagger_j c_{j+1} + c^\dagger_{j+1} c_j + 2\Delta c^\dagger_{j} c_{j}c^\dagger_{j+1} c_{j+1}\right\}\,,
\label{eq:xxzcharge}
\end{eqnarray}
obtained from (\ref{eq:xxzhamiltonianoned}) by Jordan-Wigner transformation from spin operators to spinless fermion operators \cite{jordan-1928},
\beq
S_j^+&=&c_j^\dagger e^{i\pi\sum_{i<j}c_i^\dagger c_i},\notag\\
S_j^z&=&c_j^\dagger c_j -\frac{1}{2}\,.
\eeq
In the case of a one-dimensional Hamiltonian with nearest-neighbor interactions, particle statistics is irrelevant, and alternatively the fermions can also be replaced by hardcore bosons \cite{giamarchi-book}.

The fermionic picture is especially useful in the non-interacting case ($\Delta=0$, also known as the XX limit), where (\ref{eq:xxzcharge}) is diagonal in Fourier space,
\beq
H_{XX}&=&\sum_k\epsilon_k c_k^\dagger c_k\,,\notag\\
\epsilon_k&=&-J\cos k\,.
\label{eq:xxhamiltonian}
\eeq
In the case of zero magnetization, which is of interest here, the ground state is described by the half-filled Fermi sea,
\beq
|\psi\rangle_{XX}=\prod_{-\pi/2<k\leq\pi/2}\!\!c_k^\dagger\,|0\rangle\,,
\label{eq:xxstate}
\eeq
where $|0\rangle$ is the fermionic vacuum, $c_k\ket{0}=0$. The (longitudinal) spin-spin correlation function,
\beq
G^{zz}(\ell)=\frac{1}{N}\sum_i\langle S_i^zS_{i+\ell}^z\rangle\,,
\eeq
which characterizes magnetic ordering, can be calculated exactly in the XX limit \cite{lieb-1961,wu-1966,mccoy-1967,mccoy-1967b,mccoy-1968,cheng-1967,wu-1976}. The result is a superposition of  quasi-long-range ferromagnetic and antiferromagnetic correlations, decaying by a power law,
\beq
G^{zz}(\ell)\propto\frac{1-(-1)^{\ell}}{\ell^{2}}\,.
\eeq
For finite $\D$, the extraction of correlation functions from the Bethe ansatz solution is highly non-trivial and only possible for some special cases (e.g. \cite{kitanine-2009}).

\begin{figure}[b]
\centering
\includegraphics[width=\figwb\textwidth,angle=0]{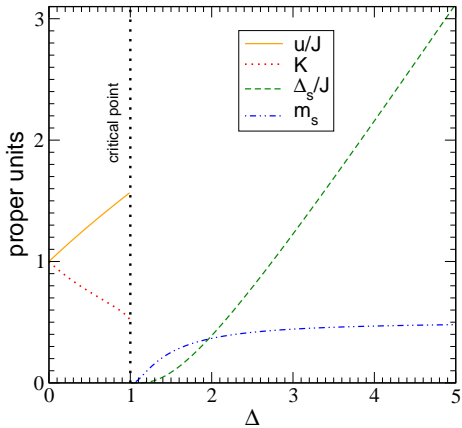}
\caption{
\label{fig:groundstateparms} The Luttinger parameter $K$, the velocity $u$, the gap $\gap$, and the staggered magnetization $\ms$ in the XXZ model, calculated by Bethe ansatz.}
\end{figure}
In order to obtain a continuum description of the Heisenberg chain, the spectrum of the non-interacting model is linearized at the Fermi points 
and the modes are separated into left- and right-movers,
\beq
H_{XX}=J\!\!\!\!\!\!\sum_{|k-\frac{\pi}{2}|\leq\frac{\L}{2J}}\!\!\!\! (k-\frac{\pi}{2})\left\{c_{R,k}^\dagger c_{R,k}-c_{L,-k}^\dagger c_{L,-k}\right\}\,.
\label{eq:xxlinearized}
\eeq
The cutoff $\L$ is of the order of the bandwidth. Starting from (\ref{eq:xxlinearized}), interactions can be included using the bosonization formalism \cite{giamarchi-book}. At the renormalization-group fixed point, which captures the long-distance properties, the Luttinger model,
\begin{equation}
H_{LL}= \frac{u}{2 \pi} \int dx \left\{ K \left( \pi \Pi(x) \right)^2 + \frac{1}{K} \left( \partial_x \phi(x) \right)^2 \right\} \ ,
\label{eq:llhamiltonian}
\end{equation}
provides the effective description for  $|\D|<1$. $\Pi(x)$ and $\phi(x)$ are conjugate bosonic fields, $[\Pi(x),\phi(x')]=i\d(x-x')$. We note that the excitations described by the Luttinger model correspond to linearly dispersed  spin waves with velocity $u$. The values of both, $u$ and the Luttinger liquid parameter $K$, can be derived from the Bethe ansatz \cite{luther-1975},
\beq
K&=&\frac{1}{2\b^2}\,,\\
u&=&\frac{J\sin(\pi(1-\b^2)) }{2(1-\b^2)}\notag,
\label{eq:llparameters}
\eeq
where $\beta$ is determined from the relation $\D=-\cos \pi\beta^2$. The functions $K(\D)$ and $u(\D)$ in the antiferromagnetic regime are plotted in Fig. \ref{fig:groundstateparms}, $K(0)=1$, $K(1)=\frac{1}{2}$, $u(0)=J$ and $u(1)=\frac{J\pi}{2}$, additionally $K(-\D)=K^{-1}(\D)$. The bosonic fields can be mapped back to the spin operators,
\beq
S^z(x) = -\frac{1}{\pi}\nabla\phi(x)+\frac{(-1)^x}{\pi\a}\cos(2\phi(x))\,,
\label{eq:magfield}
\eeq
where $\a\sim\frac{1}{\L}$. Here, the lattice spacing is set to one, so that the original sites are located at $x=i$, $i=1,\dots,N$ ($N$ being the number of lattice sites).

\begin{figure}[b]
\centering
\includegraphics[width=\figwb\textwidth,angle=0]{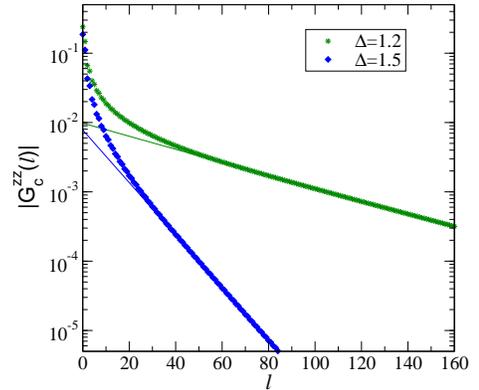}
\caption{
\label{fig:groundstate}
 Correlation function in the ground state at $\D=1.2,\,1.5$. Straight lines are exponential laws $e^{-\ell/\xi}$. While for $\D=1.5$ the inverse correlation length $\xi^{-1}=0.0873$ is very close to the value predicted by the Klein-Gordon model ($\gap/J=0.0866$), for $\D=1.2$ it is considerably larger ($\xi^{-1}=0.021$, $\gap/J=0.0048$).}
\end{figure}

For the quadratic Luttinger Hamiltonian (\ref{eq:llhamiltonian}), the correlation functions can be evaluated \cite{giamarchi-book},
\beq
G^{zz}(\ell)=C_1 \frac{1}{\ell^2}+C_2(-1)^{\ell}\left(\frac{1}{\ell}\right)^{2K}\,.
\label{eq:llcorr}
\eeq
The constants $C_1$ and $C_2$ have been calculated in Ref \cite{lukyanov-2003}. Hence, in the whole planar phase ($|\D|<1$), the correlations exhibit critical behavior and fall off algebraically. 

A different situation has to be faced for $\D\geq1$. In the renormalization-group treatment backscattering terms become important. For $\D\gtrsim1$, the sine-Gordon Hamiltonian,
\beq
H_{SG}=H_{LL}+\frac{2J\D}{(2\pi\a)^2}\int dx \cos(4\phi(x))\,,
\label{eq:sinegordon}
\eeq
is the effective model. At the isotropic point, $\D=1,K=\frac{1}{2}$, the cosine term is marginally relevant and leads to logarithmic corrections to the correlation function (\ref{eq:llcorr}). For Ising-like anisotropies, $\D>1$, $0<K<\frac{1}{2}$, the cosine term is relevant -- a spectral gap, $\gap$, opens and the phase $\phi$ becomes pinned at $0$ or $\pi/2$. Hence, $\D=1$ marks a phase transition to an antiferromagnetically ordered phase with a finite asymptotic value of the spin-spin correlations, 
\beq
G^{zz}(\ell) \mathop{=}_{\ell\rightarrow \infty} (-1)^\ell\ms^2\,.
\label{eq:orderparam}
\eeq
The two degenerate ground states, corresponding to $\phi=0$ or $\pi/2$, exhibit \textit{staggered magnetization},  $\ms$, of opposite signs,
\beq
\ms\equiv \frac{1}{N} \sum_j (-1)^j \langle S_j^z\rangle\sim \av{\cos(2\phi)}\,.
\eeq
The spectral gap as well as the staggered magnetization are continuous in all derivatives in $\D$ -- the phase transition is of Berezinskii-Kosterlitz-Thouless type \cite{berezinskii-1971,kosterlitz-1973}. In Fig. \ref{fig:groundstateparms}, $\gap$ and $\ms$ are plotted as calculated from the Bethe ansatz \cite{baxter-1973,baxter-book}. The energetically lowest excitations of the sine-Gordon model (\ref{eq:sinegordon}) are solitons and antisolitons, which create kinks to antiferromagnetic domains with negativ (solitons), respectively positiv (antisolitons), sublattice magnetization $\ms$.

For sufficiently large anisotropies, where a semiclassical approximation becomes valid, the sine-Gordon model reduces essentially  to the \textit{Klein-Gordon} Hamiltonian,
\beq
H_{KG}= H_{LL}+\gap \int dx (\phi(x))^2\,.
\label{eq:kleingordon}
\eeq
As a result of the presence of the mass term in (\ref{eq:kleingordon}), the connected correlation function,
\beq
G_c^{zz}(\ell)=\frac{1}{N}\sum_i\langle S_i^zS_{i+\ell}^z\rangle-\langle S_i^z\rangle \langle S_{i+\ell}^z\rangle\,,
\eeq
decays exponentially for large distances, 
\beq
G_c^{zz}(\ell)\sim e^{-\ell/\xi}\,,
\label{eq:expcorrelations}
\eeq
where the correlation length is given by the inverse gap,
\beq
\xi\sim\frac{J}{\gap}\,.
\label{eq:corrlength}
\eeq
In Fig. \ref{fig:groundstate} the behavior (\ref{eq:expcorrelations}) is confirmed in the gapped state of the XXZ model by numerical simulations using imaginary-time evolution of the infinite-size matrix product state (see Ref. \cite{vidal-2007} or Section \ref{sec:mps} for the description of this method). However, the relation (\ref{eq:corrlength}) is only valid for sufficiently large gaps.

In order to avoid dealing with the complicated structure of the antiferromagnetic states in the XXZ model, we introduce the {\it spin-density-wave} (SDW) state,
\beq
|\psi\rangle_\text{SDW}=\prod_{-\pi/2<k\leq\pi/2}( u_kc_k^\dagger +v_kc_{k+\pi}^\dagger)|0\rangle\,.
\label{eq:sdw}
\eeq
The coefficients of the wave function are related to the gap parameter $\gap$ by
\beq
&&v_k u_k=\frac{\gap}{2\sqrt{\epsilon_k^2+\gap^2}}\notag\,,\\ 
&&u_k^2- v_k^2=\frac{\epsilon_k}{\sqrt{\epsilon_k^2+\gap^2}}\notag\,,\\
&&u_k^2+ v_k^2=1\,.
\eeq
The correlation function calculated with the state (\ref{eq:sdw}) reproduces the exponential decay (\ref{eq:expcorrelations}) with the correlation length inversely proportional to the gap (\ref{eq:corrlength}) -- the spin-density wave provides a valid phenomenological description of antiferromagnetic states. For special values of parameters -- namely at the Luther-Emery point \cite{luther-1975} -- the spin-density wave (\ref{eq:sdw}) coincides with the exact ground state of the sine-Gordon model (\ref{eq:sinegordon}). Varying  the gap parameter $\gap$ from zero to infinity, the spin-density-wave state (\ref{eq:sdw}) links the ground state of the XXZ model at $\D=0$ (\ref{eq:xxstate}) with the N\'eel state,
\beq
|\psi\rangle_{\text{N\'eel}} = | \ua \da \ua \dots \da \ua \da \rangle\,,
\label{eq:neelapp}
\eeq
 the ground state in the limit $\Delta\rightarrow\infty$.

\section{Matrix product algorithm for time-dependent problems in the thermodynamics limit}
\label{sec:mps}

The concept of matrix product states (MPS) \cite{derrida-1993,fannes-1992,klumper-1991,klumper-1992} as a generalization of valence-bond states \cite{anderson-1987,affleck-1987,affleck-1988} has been developed parallel in time with the density matrix renormalization group (DMRG) algorithm \cite{white-1992,white-1993}. DMRG established quickly as one of the most powerful numerical approaches for solving (quasi) one-dimensional correlated many-body problems at equilibrium. Although DMRG was originally introduced as a real-space renormalization group, it can be understood as a variational optimization procedure in the space of matrix product states \cite{ostlund-1995}. This identification of DMRG with MPS is especially useful for the implementation of the ideas of DMRG in the thermodynamic limit \cite{vidal-2007,mcculloch-2008} and for time-dependent problems \cite{vidal-2003,daley-2004,white-2004,feiguin-2005}. In the following we present a formulation of a DMRG-like algorithm, which is most suitable for both time-dependent and infinite-size calculations. The procedure is identical to the infinite-size time-evolving block decimation algorithm iTEBD \cite{vidal-2007}, except that different matrices, introduced in the context of static DMRG (Ref. \cite{mcculloch-2008}), are used in order to improve the stability of the algorithm. Since neither the density matrix nor the renormalization group idea appears explicitly in this formulation, we refer to the algorithm as the matrix product state algorithm MPS or iMPS, if the infinite-size limit shall be emphasized.  Error analysis will be given for a specific case of a non-equilibrium problem in the thermodynamic limit, where we find that the behavior of the error can be considered identical to the case of the time-dependent DMRG for finite lattices \cite{gobert-2005}.

\subsection{Matrix product states}
In order to construct a MPS, we consider a one-dimensional lattice model where the Hilbert space can be separated into left and right subspaces $L^i$  and $R^{i+1}$ --  $L^i$ including $i$ as rightmost site, $i+1$ being the leftmost site of $R^{i+1}$. Generally, a wave function can be written as
\beq
\label{eq:psi}
\ket{\psi}=\sum_{\alpha\beta} |\Phi^{L^i}_\alpha\rangle(\Lambda_i)_{\alpha\beta}|\Phi^{R^{i+1}}_\beta\rangle\,,
\eeq
where $\ket{\Phi^{L^i}_\alpha}$ ($\ket{\Phi^{L^{i+1}}_\beta}$) are orthonormal basis vectors of the space $L^i$ ($R^{i+1}$). $\Lambda_i$ is called the \textit{bond center matrix} of bond $i$ and constructs the density matrix of the left and right subsystems, $\rho^{L^{i}}=\L_i^\dagger\L_i$ and $\rho^{R^{i+1}}=\L_i\L_i^\dagger$ respectively. If each site is described by a set of local basis vectors $\ket{s_i}$ of dimension $d_i$ ($s_i=0,\dots,d_i-1$), a state of the subspace can be expanded in terms of the local basis and the remaining subspace,
\beq
\label{eq:mpsexpansion}
|\Phi^{L^i}_\alpha\rangle=\sum_{\b s_i}\ket{\Phi^{L^{i-1}}_\b}(A^{s_i}_i)_{\b\a}\ket{s_i}\,.
\eeq
The orthonormality of the basis imposes on $A_i^{s}$ the left orthonormalization constraint,
\beq
\label{eq:leftorthonormal}
\sum_s A_i^{s\dagger} A_i^{s'}=\d_{ss'}\,.
\eeq
Equivalently, the state of the right subspace can be expanded by means of right orthonormalized matrices,
\beq
&&|\Phi^{R^i}_\alpha\rangle=\sum_\b\ket{s_i}(B_i^{s_i})_{\a\b}\ket{\Phi^{R^{i+1}}_\b}\,,\\
\label{eq:rightorthonormal}
&&\sum_s B_i^{s} B_i^{s'^\dagger}=\d_{ss'}\,.
\eeq
An iterative expansion of an arbitrary state $\ket{\psi}$ on a lattice of size $N$ is possible,  providing a matrix-product expression of the state,
\beq
\label{eq:mpsleft}
\ket{\psi}=\mbox{Tr}\sum_{s_1s_2\dots s_N}A_1^{s_1}A_2^{s_2}\,\dots\,A_N^{s_N}\ket{s_1s_2\dots s_N}\,.
\eeq
In the limit of $N\rightarrow\infty$, in the presence of translational symmetry, a MPS can be constructed as a periodic array of matrices. Choosing a 2-site unit cell for concreteness, i.e. $A_{i+2}=A_i$, the set of matrices $A_1^{s}$, $A_2^{s}$, $B_1^{s}$, $B_2^{s}$, $\L_1$, and $\L_2$ provides full information about the wave function. For instance, it is possible to construct the major object to be manipulated in a MPS algorithm, the \textit{two-site center matrix} (index $i$ denotes the site type, which can be either $1$ or $2$ for odd or even $i$),
\beq
\Lambda_i^{ss'}&=&A_{i}^{s}\L_i B_{i+1}^{s'},
\eeq
which can be used to decompose the wave function into the local bases of sites $i$ and $i+1$,
\beq
\ket{\psi}&=&\sum_{\a s s' \b} |\Phi^{L^{i-1}}_\alpha\rangle\ket{s}(\Lambda_i^{ss'})_{\alpha\beta}\ket{s'}|\Phi^{R^{ i+2}}_\beta\rangle\,.
\eeq
Left and right orthonormalized matrices are related via the \textit{single-site center matrix}, 
\beq
\L_i^{s}\equiv\L_{i-1}B_i^{s}=A_i^{s}\L_{i}\,,
\eeq
and can be formally mapped to each other,
\beq
\label{eq:leftright}
 B_i^{s}=\L_{i-1}^{-1}\L_i^{s}\text{ and } A_i^s=\L_i^{s}\L_{i}^{-1}\,.
\eeq
Using the single-site center matrix, the calculation of observables for a MPS representation is straightforward. For a local operator $O_i^{ss'}$ acting on site $i$, 
\beq
\langle O_i\rangle=\sum_{ss'}O_i^{ss'}(Tr\L^{\dagger s'}\L^s)\,.
\eeq
Similarly, introducing an iterative procedure, correlation functions can be calculated \cite{mcculloch-2008}.

\subsection{Schmidt decomposition}

The preceding introduction of MPS is completely general and, if infinite-dimensional matrices are allowed, any state can be formally expressed in terms of a matrix product. A class of valence-bond states \cite{affleck-1988,derrida-1993,fannes-1992,klumper-1991,klumper-1992,affleck-1987} is indeed naturally formulated in terms of MPS. Also, product states are trivially represented as MPS. Matrix product states are however especially powerful in combination with an approximative numerical algorithm, providing the optimal reduced basis set for replacing a large or possibly infinite Hilbert space. The \textit{Schmidt decomposition}, as described in the following, is the procedure which allows to select the most relevant basis states.

If only a finite number of states $m$ is supposed to be retained (in order to keep the dimension of the Hilbert space manageable for the computer), it can be shown \cite{white-1993} that a state $\ket{\tilde{\psi}}$ approximates best the targeted state $\ket{\psi}$ in the form (\ref{eq:psi}), if it is defined as the Schmidt decomposition of rank $m$ (site indices are omitted),
\beq
\label{eq:schmidt}
|\tilde{\psi}\rangle=\sum_{\alpha=1}^{m} |\tilde{\Phi}^L_\alpha\rangle\lambda_\alpha|\tilde{\Phi}^R_\alpha\rangle\,.
\eeq
The  {\it Schmidt coefficients}, $\lambda_\a$, are the dominating eigenvalues of the singular value decomposition, 
\beq
\label{eq:svd}
\Lambda_{\alpha\beta}=\sum_\g U_{\a\g}\lambda_\g V^*_{\b\g}\,,\,\,\,\lambda_1^2\geq\lambda_2^2\geq\dots
\,,
\eeq
satisfying $\sum_\alpha \lambda_\alpha^2=1$. The discarded weight,
\beq
w=\sum_{\alpha> m} \lambda_\alpha^2\,,
\eeq
corresponds to the mismatch, $|\ket{\tilde{\psi}}-\ket{\psi}|=w$, introduced by this truncation procedure. The new basis is given in terms of the \textit{Schmidt states},
\beq
\ket{\tilde{\Phi}^L_\g}&=&\sum_{\a}U_{\a\g}\ket{\Phi^L_{\a}}\,,\notag\\
\ket{\tilde{\Phi}^R_\g}&=&\sum_{\a}V^*_{\a\g}\ket{\Phi^R_{\a}}\,.
\label{eq:schmidtstates}
\eeq
In practice it is useful to set only an upper bound for $m$ (rather than fixing a definite value) and instead define a threshold $\e$ such that only states for which $\lambda_\a^2\geq\e$ are retained. The applicability of this truncation procedure to a physical state depends on the characteristics of the Schmidt values or, equivalently, the spectrum of the density matrix. The more slowly the values $\lambda_\a$ decay, the larger must be the number of retained states. A generic expression for the spectrum of the density matrix has been obtained for a critical theory \cite{calabrese-2008}, for practical purposes  it is however sufficient to consider the entanglement properties of the system to get the order of the necessary number of retained states. For instance, one can consider the \textit{entanglement entropy}, 
\beq
S=Tr (\rho^{L,R} \log_2 \rho^{L,R})=\sum_\a \lambda_\a^2\log_2\lambda_\a^2\,.
\eeq
The fact that in one-dimensional equilibrium states the entanglement entropy exhibits logarithmic dependence on the typical length scale $\xi$ of the state \cite{calabrese-2004} ($\xi$ corresponds to the correlation length or, at criticality, to the size of the system) guarantees an accurate description of a large class of wave functions using a finite number of states $m\propto \xi$. Away from equilibrium, however, the entanglement generally grows linearly in time \cite{calabrese-2005} and a potentially exponential growth of $m$ with time restricts the applicability of a MPS to short times.

For a wave function represented at bond $i$ by the two-site center matrix $\L_i^{ss'}$, the Schmidt decomposition reads as follows: Replacing  in Eq. (\ref{eq:psi}) the contracted two-site center matrix with the bond center matrix,
\beq
\L_{d_is+\a,d_is'+\b}=(\L^{ss'}_i)_{\a\b}\,,
\eeq
 the Schmidt decomposition can be carried out as presented above. The matrices are updated retaining $m$ Schmidt states (\ref{eq:schmidtstates}),
\beq
&&(A_i^s)_{\a\b}\rightarrow U_{d_1+\a,\b} \notag\\
&&(B_{i+1}^s)_{\a\b} \rightarrow V^*_{\a,d_2s_2+\b}\\
&&(\L_{i})_{\a\b}\rightarrow\d_{\a\b}\lambda_\a\,,\notag
\eeq
with $\L_{i+1}$ remaining unchanged.

For the evaluation of correlation functions, additionally $A_{i+1}$ or $B_{i}$ are needed. Although the effect of loss of orthogonality is spurious when applying the direct inverse (\ref{eq:leftright}) as in the original iTEBD algorithm \cite{vidal-2007}, especially in the case of real-time evolution, a procedure for recovering both left and right orthonormalized representations is  needed for stabilizing the algorithm. The left (right) \textit{rotation} of the matrices does the job. Starting from a single-site center matrix, $\Lambda_i^{s}=\L_{i-1} B^{s}_{i}$, the left orthonormalized matrix and the rotated center matrix can be extracted from the singular value decomposition (\ref{eq:svd}) of the re-indexed matrix, $\L_{\a+ds,\b}=(\L^{s})_{\a\b}$,
\beq
(A_i^{s})_{\a\b}=U_{\a +ds,\b}\,,\,\,(\L^R_{i})_{\a\b}=\lambda_\a V^*_{\b\a}\,.
\eeq
An iterative application of this procedure moves the center matrix through the lattice and brings all matrices into left orthonormal form. An analogous left-moving iteration brings the matrices into the right orthonormalized form. In the periodic iMPS a problem arises when the right-moving center matrix reaches the edge of the unit cell. $\L^R_{i}$ does not in general coincide with the former $\L_{i+1}$ and repeating the iterations through the unit cell further changes the MPS. There exist however schemes which solve this problem by introducing an additional transformation, after which the \textit{transfer operator} $\L^R_{i}\L_{i+1}^{-1}$ becomes equal to identity (see Refs. \cite{orus-2008,mcculloch-2008} for detailed descriptions).

\begin{figure*}[t]
\includegraphics[width=0.95\textwidth,angle=0]{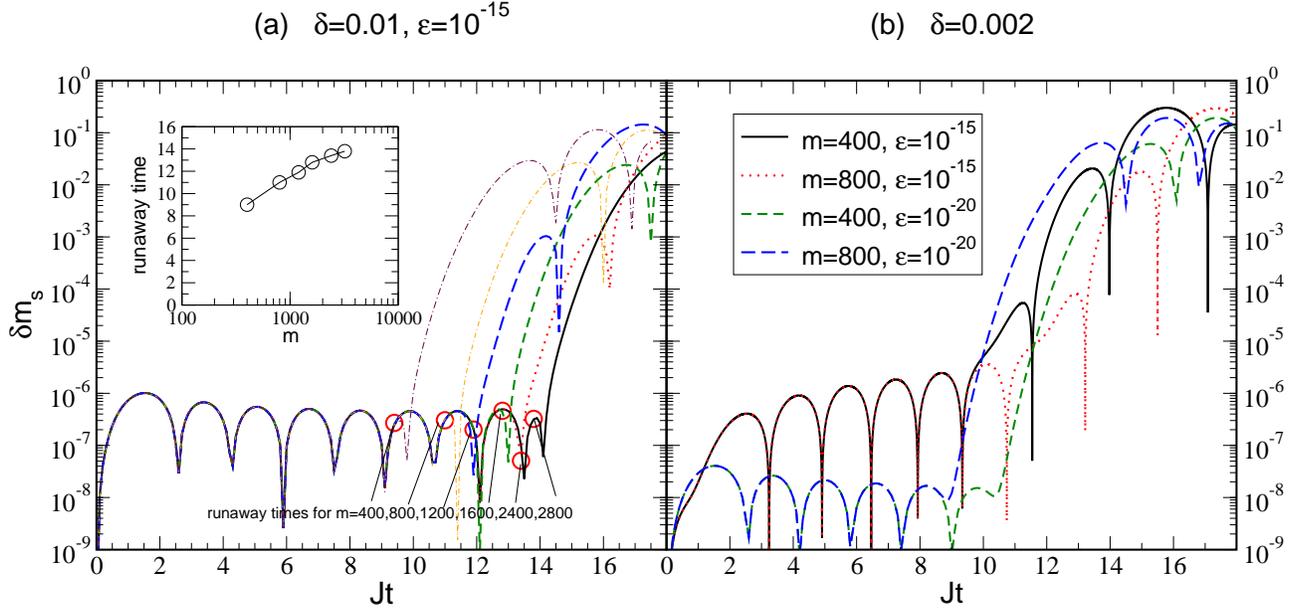}
\caption{\label{fig:error}The error in an iMPS compared to the exact solution. (a) Trotter slicing $\d=0.01$, threshold $\e=10^{-15}$. Inset: runaway times as a function of $m$. (b) Comparison of errors for two different $\e=10^{-15}$, $10^{-20}$, for a smaller slicing $\d=0.002$.}
\end{figure*}

\subsection{Suzuki-Trotter decomposition}
In order to calculate the time evolution of a MPS, $\ket{\psi(t)}=e^{-iHt}\ket{\psi_0}$, it is suitable to approximate the evolution operator, $e^{-iHt}$, by a Suzuki-Trotter decomposition. This is possible if the global operator $H$ contains only nearest-neighbor bond terms $H=\sum_i H_{i,i+1}$ (e.g. the Heisenberg chain with $H_{i,i+1}={\bf S}_i{\bf S}_{i+1}$). $H$ can then be decomposed into even and odd parts $H=H_1+H_2$,
\beq
H_1=\sum_j H_{2j,2j+1}\,,\,\,H_2=\sum_jH_{2j+1,2j+2}\,.
\eeq
The Suzuki-Trotter decomposition can be regarded as the first-order expansion of the evolution operator using the Baker-Hausdorff formula \cite{suzuki-1990},
\beq
e^{-iHt}=(e^{-iH_2\d}e^{-iH_1\d})^n+O(\d^2n)\,,\,\,n\d=t\,.
\eeq
This approximation is improved in a second-order expansion,
\beq
e^{-iHt}=(e^{-iH_1\d/2}e^{-iH_2\d}e^{-iH_1\d/2})^n+O(\d^3n)\,,
\eeq
or, if higher accuracy is desired, using third- or higher-order expansions \cite{mclachlan-1995}.

Since the components of the even (odd) part commute with each other, 
\beq
[H_{2j,2j+1},H_{2j',2j'+1}]=[H_{2i+1,2i+2},H_{2i'+1,2i'+2}]=0\,,\notag
\eeq
within the first-order Suzuki-Trotter decomposition the evolution operator can be broken down to a product of nearest-neighbor operators,
\beq
\label{eq:exp1}
e^{-iHt}&\!\!\!\approx&\!\!\!\left (\prod_i e^{-i{H_{2i+1,2i+2}}\d}\prod_j e^{-i{H_{2j,2j+1}}\d}\right)^n\!\!.
\eeq
Equivalent expressions hold for higher-order decompositions.

The Suzuki-Trotter decomposition can also be used for calculating the ground state $\ket{\psi}$ of a Hamiltonian using \textit{imaginary}-time evolution,
\beq
\ket{\psi}\mathop{=}_{\tau \rightarrow\infty} \frac{e^{-\tau H}\ket{\psi_0}}{|e^{-\tau H}\ket{\psi_0}|}\,,
\eeq
where $\ket{\psi_0}$ is some random initial state. In order to get reliable results from this procedure, the Trotter slicing has to be reduced carefully during the imaginary-time evolution \cite{vidal-2007}.

\subsection{Update of an iMPS}
We consider now the application of a single factor of (\ref{eq:exp1}) onto a MPS. For example, for the odd bond operator, $U=e^{-iH_{1,2}\d}$, we have
\beq
\tilde{\L}^{s'_1s'_2}=\sum_{s_1s_2}U_{s'_1s'_2;s_1s_2}\L_1^{s'_1s'_2}\,.
\eeq
After a subsequent singular value decomposition, retaining a finite number of states, the matrices can be updated,
\beq
\label{eq:mpsupdate}
\tilde{\L}^{ss'}=A_1^{s}\L_1 B_2^{s'}.
\eeq
In the case of an iMPS of periodicity 2, the effect of the remaining factors of $e^{-i{H_1}\d}$ on the other odd bonds is identical. Hence, the update (\ref{eq:mpsupdate}) corresponds to the action of the operator $e^{-iH_{1}\d }$ on the whole, infinitely extended wave function. After the update one can recalculate $B_1^{s}$, $A_2^{s}$  by means of left- and right-moving iterations. More straightforward for preparing the application of the odd bond operator $e^{-iH_{2}\d}$ is the direct construction of the center matrix,
\beq
\L^{ss'}=\L_1 B_2^{s}\L_2^{-1}A_1^{s'}\L_1\,.
\eeq
Since the inverse of $\L_2$ is required, a finite threshold $\e$ is necessary to guarantee the stability of this operation.

\subsection{Error analysis}

As an application of the iMPS method to a non-equilibrium problem, we study the quench problem in the XXZ chain, $\ket{\psi(t)}=e^{-iHt}\ket{\psi_0}$, where $\ket{\psi_0}$ is the ground state of the XXZ Hamiltonian at a given value $\D=\D_0$, and $H$ is characterized by an anisotropy parameter $\D$.
In this case the local basis consists of a spin-up and a spin-down states $\{\ket{\downarrow},\ket{\uparrow}\}$. This quench problem is analyzed in detail in section \ref{sec:xxz}.

\begin{figure}[ht]
\centering
\includegraphics[width=\figwa\textwidth,angle=0]{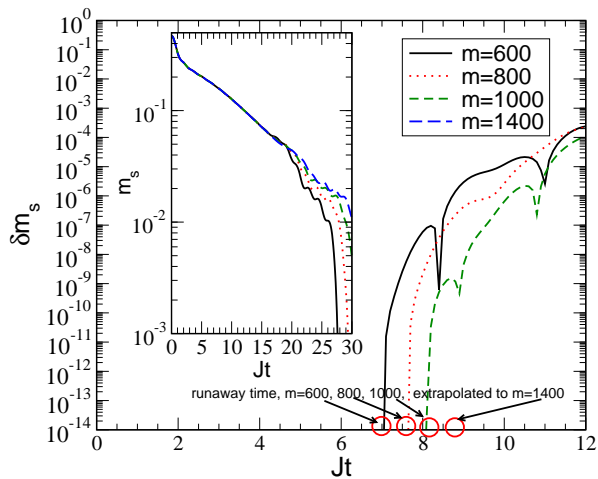}
\caption{\label{fig:error2}The absolute error (\ref{eq:error2}) in an iMPS simulation for $\d=0.005$, $\e=10^{-17}$ for different values of $m$. Inset: the corresponding dynamics of $\ms(t)$.}
\end{figure}

First we study the case where $\ket{\psi_0}$ is the N\'eel state which has a trivial iMPS representation with $m=1$, $A_1^s=B_1^s=\d_{s\uparrow}$, $A_2^s=B_2^s=\d_{s\downarrow}$. Since the $z$-projection of the total spin ($S^z_{tot}$) is conserved, the MPS can be resolved by this quantum number \cite{mcculloch-2008,mcculloch-2007}. The resulting speedup is about an order of magnitude in comparison with a simulation which exploits no symmetry. In the limit $\D=0$ the numerical results can be checked against the exact solution (see section \ref{sec:xx}). In Fig. \ref{fig:error} the absolute deviation from the exact result,
\beq
\d \ms(t)=|\ms(t)-J_0(2Jt)/2|\,,
\label{eq:error1}
\eeq
is plotted for different values of the number of retained states $m$, the threshold $\e$ and the Trotter slicing 
$\d$. The evolution of the error can be clearly divided into two regimes by introducing the runaway time $t_{runaway}$: For $t<t_{runaway}$ there is a small error which does not depend on the value of $m$. In this case the error is dominated by the Trotter error, which grows at most linearly in time. However, for $t>t_{runaway}$ the error starts growing nearly exponentially. The approximately logarithmic dependence of $t_{runaway}$ on $m$ (Fig. \ref{fig:error}(a), inset) is in agreement with the linear growth of the entanglement entropy in the non-equilibrium problem -- $t_{runaway}$ can be understood as the point where the chosen finite number of retained states is no more sufficient to represent the entanglement in the state. We note, however, that a strict relation between entanglement entropy and $t_{runaway}$ can not be rigorously established \cite{gobert-2005}.

In order to reduce the Trotter error dominating at $t<t_{runaway}$, one may choose smaller values of $\d$. The threshold $\e$ has to be decreased as well. Otherwise, due to the increased number of updates, errors associated with the discarded weight at each step may accumulate. In Fig. \ref{fig:error}(b) we plot two cases with $\e=10^{-15}$ and $10^{-20}$. In general, it is sufficient to reduce the threshold proportional to the Trotter slicing $\e\propto\d$t.

If the Trotter slicing is chosen so that the resulting error is of the order of the accuracy goal, $t_{runaway}$ then sets the time window for the validity of the numerical results (in Fig. \ref{fig:error} the accuracy goal in the absolute error is about $10^{-6}$). We note that such behavior of the error in this time-dependent infinite-size MPS algorithm is identical to that of the finite-size DMRG algorithm \cite{daley-2004}.

If the exact solution is not known, $t_{runaway}$ can nevertheless be determined by comparing curves from calculations with slightly different $m$. $t_{runaway}$ is the point where the difference between them starts to grow significantly. Fig. \ref{fig:error2} illustrates this procedure with the results for a quench in the XXZ chain from $\Delta_0=4$  to $\Delta=2$, with $\d=0.005$ and $\e=10^{-17}$ (see also section \ref{sec:xxz}). Comparing  the difference for various $m$,
\beq
\d \ms(t)=|\ms(t)-\ms(t)^{m=1400}|\,,
\label{eq:error2}
\eeq 
where $\ms(t)^{m=1400}$ is the result for 1400 retained states, we find a behavior identical to the exactly solvable case of the XX chain -- the curves for $m<1400$ overlap completely with the one for $m=1400$ up to $t<t_{runaway}$ and a difference can only be seen for $t>t_{runaway}$. For $m=1400$,  $t_{runaway}$ is estimated in Fig. \ref{fig:error2} by extrapolation of the values for $m=600,800,1000$. Again, the accuracy of the results for $t<t_{runaway}$ are dominated by the Suzuki-Trotter error which has to be estimated separately (here it is of the order of $10^{-7}$). In practice it is not mandatory to abort the calculation at the runaway time -- from the rough behavior of $\d \ms(t)$ one can estimate that even for $t\leq 10$, the absolute error of the curve for $m=1400$ is still of the order of $10^{-6}$.

The presented error analysis has been carried out for a local parameter in a specific setup. As long as non-equilibrium dynamics is concerned, this behavior is completely generic, although the runaway time and the Suzuki-Trotter error have to be determined for each case. Also, the error may depend on the observable under consideration -- long-distance correlation functions may exhibit shorter runaway times than local observables. We would like to emphasize that error control, which imposes criteria on the \textit{wave function} \cite{garcia-2006}, is in general too strict, and the presented observable-based approach can considerably extend the accessible time window.

\end{document}